\documentclass[aps,prc,eqsecnum,preprint,tightenlines,nofootinbib]{revtex4-2}
\usepackage{graphicx,latexsym}
\usepackage{centernot}

\def\bea{\begin{eqnarray}}
\def\eea{\end{eqnarray}}
\def\be{\begin{equation}}
\def\ee{\end{equation}}

\newcommand{\fsl}[1]{{\centernot{#1}}}
\newcommand{\mQ}{\left(\frac{m}{Q}\right)}

\begin{document}

\title{Reduced nuclear helicity amplitudes for deuteron electrodisintegration
and other processes}

\author{J. Flores}
\author{S. S. Chabysheva}
\affiliation{Department of Physics, University of Idaho, Moscow ID 83844 USA}
\author{J. R. Hiller}
\affiliation{Department of Physics, University of Idaho, Moscow ID 83844 USA}
\affiliation{Department of Physics and Astronomy,
University of Minnesota-Duluth,
Duluth, Minnesota 55812 USA}

\date{\today}

\begin{abstract}
We extend the original idea of reduced nuclear amplitudes to capture
individual helicity amplitudes and discuss various applications to
exclusive processes involving the deuteron.  Specifically, we consider deuteron
form factors, structure functions, tensor polarization observables,
photodisintegration, and electrodisintegration.  The
basic premise is that nuclear processes at high momentum transfer
can be approximated by tree graphs for point-like nucleons supplemented
by empirical form factors for each nucleon.  The latter represent the
internal structure of the nucleon, and incorporate nonperturbative
physics, which can allow for early onset of scaling behavior.  The nucleon
form factors are evaluated at the net momentum transfer experienced by
the given nucleon, with use of $G_E$ for a no-flip contribution
and $G_M$ for a helicity-flip contribution.  Results are compared with 
data where available.  The deuteron photodisintegration asymmetry 
$\Sigma$ is obtained with a value of $\Sigma(90^\circ)\simeq -0.06$, 
which is much closer to experiment than the value of -1 originally expected.
The method also provides an estimate of the
momentum transfer values required for scaling onset.  We find that
the deuteron structure function $B$ is a good place to look, above 
momentum transfers of 10 GeV$^2$.

\end{abstract}

\maketitle

\section{Introduction}
\label{sec:Introduction}

With the advent of the upgraded electron accelerator at the Thomas
Jefferson National Accelerator Facility, scattering experiments with
polarized beams and targets at high energy and high momentum transfer
become possible.  In the regime of high momentum transfer to all
relevant nucleons, quantum chromodynamics (QCD) implies that
the internal structure of every nucleon is important.  Until
{\em ab initio} QCD (lattice) calculations for nuclear scattering processes
are available for more than very simple processes, one 
is led to consider models that can represent the basic physics.

One such approach is the reduced nuclear amplitude (RNA) analysis
pioneered by Brodsky and Chertok~\cite{BrodskyChertok}.  In addition
to their application to a generic deuteron form factor, the approach
has been applied to deuteron disintegration~\cite{RNA}, pion 
photoproduction~\cite{pion}, and photodisintegration of $^3$He~\cite{3He}.
As originally developed, a nuclear process was modeled as a tree-level
amplitude multiplied by a generic form factor for each nucleon, with
each form factor evaluated at the net momentum transferred to that nucleon.
In order to model the behavior of polarization 
observables~\cite{Jeschonnek1,Jeschonnek2a,Jeschonnek2b,Jeschonnek3,%
Laget,Atti,Sargsian2009,Arenhovel,Gakh,Raskin,Dmitrasinovic}, we
extend this approach to a reduced nuclear helicity amplitude (RNHA) method
to combine a tree-level helicity amplitude for point-like nucleons 
with the appropriate form factor for each
nucleon.  When the nucleon does (not) flip its helicity, we
use the electric (magnetic) form factor $G_{EN}$ ($G_{MN}$).
As a check on the procedure, virtual photon absorption by a single 
nucleon in the RNHA approach is consistent with
the definitions of $G_{EN}$ and $G_{MN}$.

A caveat in applications of the RNA approach is that the normalization is
not determined by the model and is fixed to data at infinite momentum
transfer by the coefficient of the leading power-law behavior.  
This means that the normalization cannot
be determined in practice; fitting to a data point at some intermediate
kinematics will give the wrong normalization and the wrong magnitude
at higher momentum transfer.  Instead, ratios need to be considered,
so that the normalization becomes irrelevant.

\begin{figure}[ht]
\vspace{0.2in}
\centerline{\includegraphics[width=10cm]{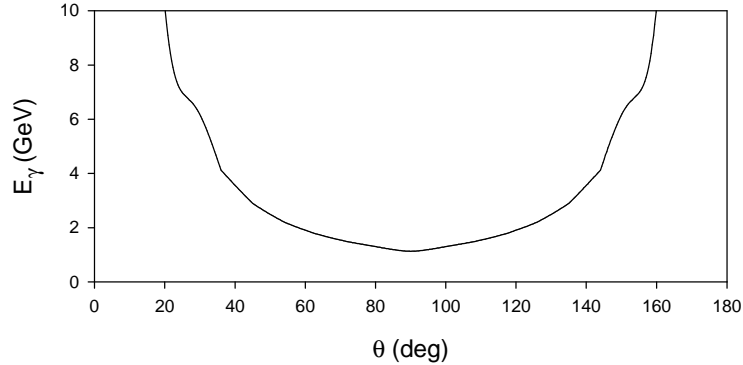}}
\caption{\label{fig:limit} Angular dependence of the scale
for large momentum transfer in deuteron photodisintegration.
}
\end{figure}

The primary criterion for the asymptotic region is in the 
momentum transfer to each nucleon.  For every nucleon in
the process, the momentum transfer must be above some
common threshold, which is at least 1 GeV$^2$.  For example,
for deuteron photodisintegration, the momentum transfer to
a nucleon is $-t_N =-(p_N-p/2)^2$, where $p_N$ is the final
four-momentum of the nucleon and $p$ is the initial deuteron momentum.
When expressed in terms of the photon energy $E_\gamma$
and the final nucleon angle $\theta$, the constraint to
be above 1 GeV$^2$ becomes~\cite{CHHreview}
\be
m_N E_\gamma\left[1-
\sqrt{\frac{E_\gamma}
   {m_N+E_\gamma}}|\cos\theta|\right] \geq 1\,\mbox{GeV}^2.
\ee
This relationship is illustrated in Fig.~\ref{fig:limit}.
Notice that away from 90$^\circ$, the lower limit is quite high.
For electrodisintegration, only the most recent data~\cite{Yero,YeroThesis}
begins to reach this threshold.

Here we will focus on deuteron processes, including photodisintegration
and electrodisintegration.  For recent reviews of deuteron studies
at high momentum transfer, see \cite{BoeglinSargsian,GilmanGross,CHHreview}.
Elastic electron scattering data at high momentum transfer is presented 
in \cite{Arnoldetal,Bosted,AbbotA,Alexa,AbbotT}.
Recent photodisintegration data can be found in \cite{Bochna,Schulte1,Schulte2,Mirazita},
and for electrodisintegration data, in \cite{Kasdorp,Ulmer,Boeglin,Egiyan,Yero,YeroThesis}.
Other analyses of deuteron processes include 
hidden-color contributions to deuteron form factors~\cite{BrodskyJiLepage},
the hard rescattering mechanism~\cite{Sargsian},
quark-gluon strings~\cite{Grishina},
the Moscow NN potential~\cite{Knyr},
and AdS/QCD models~\cite{Huseynova,Gutsche}.
	
One recent experiment~\cite{Yero} used the 10.6 GeV electron beam
at JLab and the Hall C spectrometers to measure electron
scattering from a liquid deuterium target.  The final electron
and the proton were detected, with the kinematics restricted
to the exclusive process $ed\rightarrow e'pn$.  One spectrometer
measured the final electron at a nominal 12.2$^\circ$ degrees 
from the beam direction, with a momentum of 8.5-9.1 GeV such 
that the recorded events had a distribution of momentum transfer 
squared reaching 5 GeV$^2$.  The events studied were taken from
a bin of 4.5$\pm$0.5 GeV$^2$ in the tail of the distribution; however,
the nominal transfer was 4.2 GeV$^2$, because the majority of the 
events were in the lower half of the bin.

A second spectrometer measured the proton momentum at a range
of angles to the beam direction, tuned to select
events where the (missing) neutron had an angle relative to the 
direction of the momentum transfer that fell within a chosen 
bin.  In the one-photon exchange
approximation, which we assume, the momentum transferred is,
of course, the photon momentum.  The published neutron angles 
are binned at 35$^\circ$, 45$^\circ$ and 75$^\circ$, with the first two
selected to minimize final-state interactions.  For our purposes,
the importance of these two angles is that the momentum
transferred to the neutron reaches 1 GeV in a zero-binding
approximation, so that, rather than focus on the internal structure
of the deuteron, we can consider the response to a large momentum transfer
to all the nucleons involved and we can see that experiments may be
approaching the threshold where our model can be applied.

The RNHA model is constructed
in detail in Sec.~\ref{sec:construction} for two-nucleon processes.
In the remainder of the paper, we consider various processes
for the deuteron.  In Sec.~\ref{sec:elasticscattering}, the form
factors,\footnote{For discussion
specifically in terms of perturbative QCD, see \protect\cite{BrodskyJiLepage}.} 
structure functions, and tensor polarization observables of 
elastic electron scattering from the deuteron 
are obtained.  Photodisintegration
and electrodisintegration are analyzed in Secs.~\ref{sec:photodis}
and \ref{sec:electrodis}. Within the zero-binding approximation,
elastic scattering and
photodisintegration live at edges of the kinematic range of
electrodisintegration and are essentially special cases that
provide introductory examples.  Section~\ref{sec:summary} contains
a summary of the results and suggestions for additional applications.
Many details of the electrodisintegration helicity amplitudes 
are left to an appendix.

\section{Construction of the model}
\label{sec:construction}

\begin{figure}[ht]
\vspace{0.2in}
\begin{center}
\begin{tabular}{cc}
\includegraphics[width=8cm]{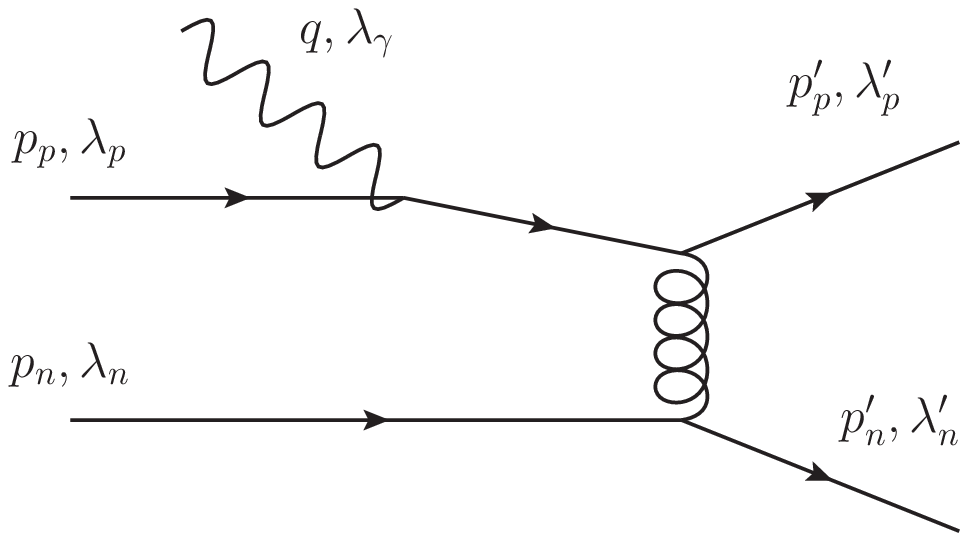} &
\includegraphics[width=8cm]{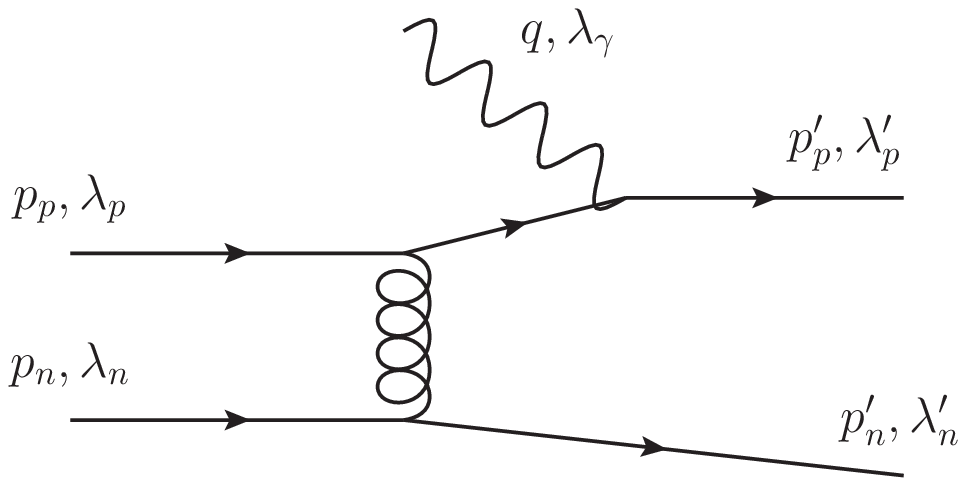} \\
(a) & (b) \\
\includegraphics[width=8cm]{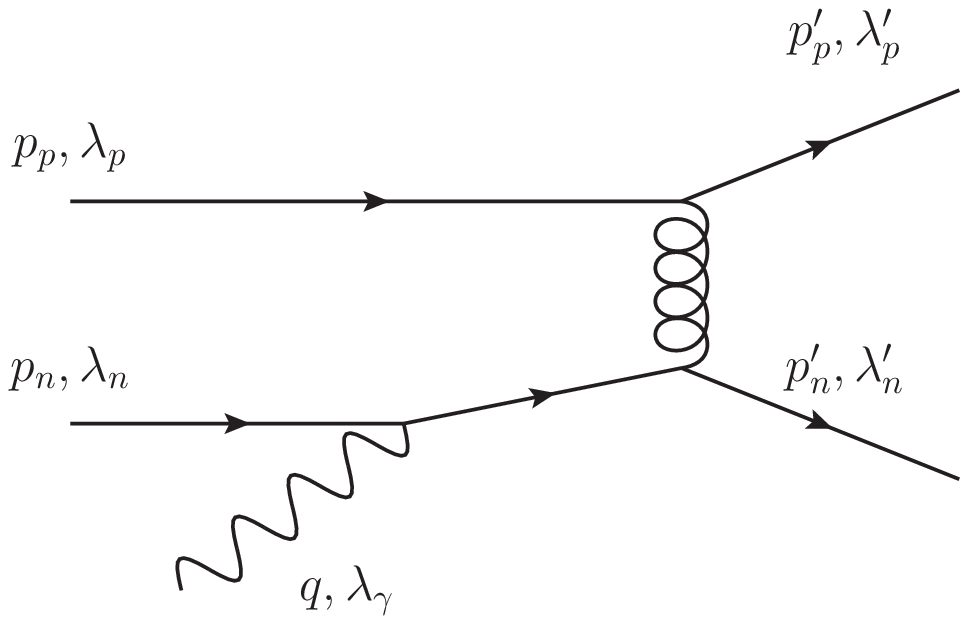} &
\includegraphics[width=8cm]{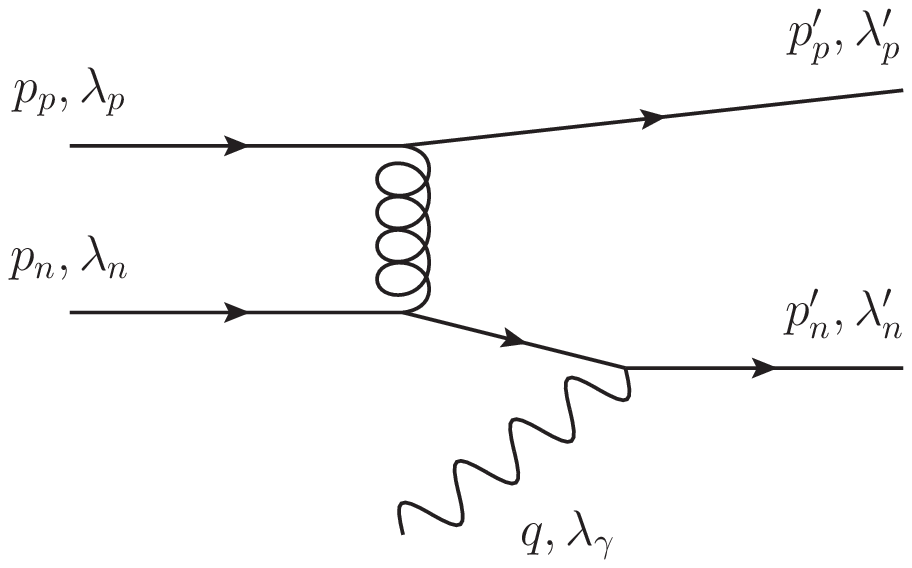} \\
(c) & (d)
\end{tabular}
\end{center}
\caption{\label{fig:diagrams} Tree graphs for deuteron
processes that absorb a photon of momentum $q$ and helicity $\lambda_\gamma$.  
The initial (final) nucleon momentum and helicity are $p_N$ ($p'_N$) and $\lambda_N$
($\lambda'_N$), with $N=p$ or $n$.  The two nucleons exchange momentum via a
vector particle.  The four diagrams differ in the nature of the photon-absorbing
nucleon and the order of this absorption and momentum transfer between nucleons.
}
\end{figure}

The basic process for a two-nucleon system to absorb a photon
and exchange momentum between the nucleons is illustrated in
Fig.~\ref{fig:diagrams}.  These diagrams are modeled on the 
primitive process of $\gamma^*ff\rightarrow ff$, with $f$ representing
a point-like nucleon.\footnote{In \cite{RNA}, the primitive
process was $\gamma^* q\bar{q}\rightarrow q\bar{q}$, with
$q$ corresponding to a point-like proton and $\bar{q}$ to a point-like
neutron, and direct interaction of the photon with the neutron was 
neglected.  Here we amend and extend this, to retain
information about helicity states of the fermions and include
photon absorption by the neutron.}
The structure of each nucleon is then introduced by combining
the Feynman amplitude for each diagram with the appropriate
form factor for each nucleon, evaluated at the net momentum
transfer for that nucleon.  For a deuteron process in the zero-binding limit, 
the initial nucleons share the initial deuteron momentum $p$
equally, so that $p_p=p_n=p/2$.  We also neglect the nucleon mass
difference, setting $m_p=m_n\equiv m$.  The distinction between 
different photon-absorption processes is then in the 
nature of the photon, being either real or virtual, and
in the outcome for the final nucleons, bound as a deuteron
or not.

The tree-level amplitudes for the four diagrams in Fig.~\ref{fig:diagrams}
are
\bea
M_a^\nu(\lambda'_p,\lambda'_n,\lambda_p,\lambda_n)
  &=& A_p^{\mu\nu}(p/2+q;\lambda'_p,\lambda_p) \frac{1}{(p'_n-p/2)^2} B_{n\mu}(\lambda'_n,\lambda_n), \\
M_b^\nu(\lambda'_p,\lambda'_n,\lambda_p,\lambda_n)
  &=& A_p^{\nu\mu}(p'_p-q;\lambda'_p,\lambda_p) \frac{1}{(p'_n-p/2)^2} B_{n\mu}(\lambda'_n,\lambda_n), \\
M_c^\nu(\lambda'_p,\lambda'_n,\lambda_p,\lambda_n)
  &=& A_n^{\mu\nu}(p/2+q;\lambda'_n,\lambda_n) \frac{1}{(p'_p-p/2)^2} B_{p\mu}(\lambda'_p,\lambda_p), \\
M_d^\nu(\lambda'_p,\lambda'_n,\lambda_p,\lambda_n)
  &=& A_n^{\nu\mu}(p'_n-q;\lambda'_n,\lambda_n) \frac{1}{(p'_p-p/2)^2} B_{p\mu}(\lambda'_p,\lambda_p),
\eea
where
\be
A_N^{\mu\nu}(p;\lambda'_N,\lambda_N)=\bar{u}'_N\gamma^\mu\frac{\fsl{p}+m}{p^2-m^2}\gamma^\nu u_N,\;\;
B_N^\mu(\lambda'_N,\lambda_N)=\bar{u}'_N\gamma^\mu u_N,
\ee
with $u_N$ ($\bar{u}'_N$) the initial (final) spinor for the nucleon $N$ with helicity $\lambda_N$ 
($\lambda'_N$).  The sub-amplitude $A_N$
represents the fermion line that absorbs the photon, and $B_N$ represents the other fermion line.
Calculation of these sub-amplitudes can be checked against the trace theorem for sums over
helicities:
\bea
\sum_{\lambda_N,\lambda'_N} A_N^{\nu'\mu'*}(p;\lambda'_N,\lambda_N) A_N^{\mu\nu}(p;\lambda'_N,\lambda_N)
&=&{\rm Tr}\left[\gamma^{\nu'}\frac{\fsl{p}+m}{p^2-m^2}\gamma^{\mu'}(\fsl{p}'_N+m)
                    \gamma^\mu\frac{\fsl{p}+m}{p^2-m^2}\gamma^\nu(\fsl{p}_N+m)\right], \nonumber \\ \\
\sum_{\lambda_N,\lambda'_N} B_N^{\mu'*}(\lambda'_N,\lambda_N) B_N^{\mu}(\lambda'_N,\lambda_N)
&=&{\rm Tr}\left[\gamma^{\mu'}(\fsl{p}'_N+m)\gamma^\mu(\fsl{p}_N+m)\right].
\eea

The full amplitude is constructed from the $M_X$ by combining them
with form factors for each nucleon.  For a deuteron with initial
helicity $\lambda_d$, we have
\be  \label{eq:Mnu}
M^\nu(\lambda'_p,\lambda'_n,\lambda_d)=\sum_{\lambda_p,\lambda_n}
  C_{\lambda_p\lambda_n}^{\lambda_d}
	\left[\sum_{X=a,b,c,d} M_X^\nu(\lambda'_p,\lambda'_n,\lambda_p,\lambda_n)\right]
		G_{p\lambda'_p\lambda_p}(Q^2_p) G_{n\lambda'_n\lambda_n}(Q^2_n),
\ee
where $Q^2_N=-(p'_N-p_N)^2$,
\be
C_{\lambda_p\lambda_n}^{\lambda_d}=\left\{\begin{array}{ll}
    \delta_{\lambda_p\pm\frac12}\delta_{\lambda_n\pm\frac12}, & \lambda_d=\pm 1 \\
		\frac{1}{\sqrt{2}}\left(\delta_{\lambda_p\frac12}\delta_{\lambda_n-\frac12}+
		                        \delta_{\lambda_p-\frac12}\delta_{\lambda_n+\frac12}\right), &
										\lambda_d=0, \end{array}\right.
\ee
and
\be
G_{N\lambda'\lambda}=\left\{\begin{array}{ll} G_{EN}, & \lambda'=\lambda \\
                                              G_{MN}, & \lambda'=-\lambda. \end{array}\right.
\ee
The form factors $G_{EN}$ and $G_{MN}$ represent the internal structure of the nucleons.
They can be represented by data or empirical fits.  For simplicity,
we use the fits~\cite{FFratios}
\be
G_{Ep} \simeq \left(1+\frac{Q_N^2}{m_0^2}\right)^{-2},\;\;
G_{Mp} \simeq \mu_p G_{Ep},\;\;
G_{Mn} \simeq \mu_n G_{Ep},\;\;
G_{En} \simeq -\frac{\mu_n \tau}{1+5.6\tau}G_{Ep},
\ee
where $m_0^2=0.71$ GeV$^2$, $\tau=\frac{Q_N^2}{4m^2}$, 
$\mu_p=2.79$, and $\mu_n=-1.91$.  To limit the analysis
to a single mass scale, we take the parameter $m_0$ to
be proportional to the nuclear mass, with $m_0^2=0.80\,m^2$.
We do not attempt to compute or assign an overall normalization
to $M^\nu$, and the running of the strong coupling constant is not included.

The initial nucleon spinor, for a deuteron traveling along the negative $z$ direction,
is~\cite{Scadron}
\be  \label{eq:initialspinors}
u_N=\frac{\fsl{p}/2+m}{\sqrt{E_d/2+m}}
  \left(\begin{array}{c} \phi^{(\lambda_N)}(-\hat z) \\ 0 \end{array}\right),
\ee
with
\be
\phi^{(1/2)}(-\hat z)=\left(\begin{array}{c} 0 \\ 1 \end{array}\right), \;\;
\phi^{(-1/2)}(-\hat z)=\left(\begin{array}{c} 1 \\ 0 \end{array}\right).
\ee
The final nucleon spinor is
\be
u'_N=\frac{\fsl{p}'_N+m}{\sqrt{E'_N+m}}
  \left(\begin{array}{c} \phi^{(\lambda'_N)}(\hat p'_N) \\ 0 \end{array}\right), 
\ee
with
\be
\phi^{(1/2)}(\hat p'_N)=
   \left(\begin{array}{c} \cos(\theta_N/2) \\ e^{i\phi_N}\sin(\theta_N/2) \end{array}\right), \;\;
\phi^{(-1/2)}(\hat p'_N)=
   \left(\begin{array}{c} -e^{-i\phi_N}\sin(\theta_N/2) \\ \cos(\theta_N/2) \end{array}\right),
\ee
where $\theta_N$ and $\phi_N$ are the polar and azimuthal angles of the
outgoing momentum of the particular nucleon.

As discussed in the Introduction,
the overall normalization of the RNHA amplitude is unknown.
For comparison with data, we consider quantities which 
are themselves ratios or a ratio of the model to data.

\section{Elastic electron scattering}
\label{sec:elasticscattering}

\subsection{Form factors} \label{sec:formfactors}

\begin{figure}[ht]
\vspace{0.2in}
\centerline{\includegraphics[width=10cm]{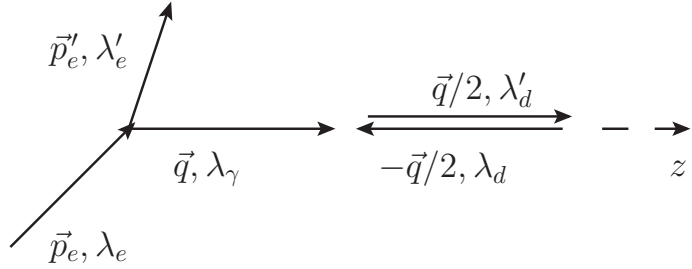}}
\caption{\label{fig:Breit} Kinematics for elastic electron-deuteron scattering
in the Breit frame.  The photon travels along the positive $z$ direction,
and the deuteron comes from the right, along the negative $z$ direction.
}
\end{figure}

The three deuteron form factors, $G_C$, $G_M$, and $G_Q$, are readily obtained from
the hadronic helicity amplitudes of elastic electron-deuteron
scattering in the Breit frame~\cite{Arnold}.  The kinematics are shown
in Fig.~\ref{fig:Breit}.  The photon four-momentum is
$q=(0,0,0,q_z)$ and the initial (final) deuteron four-momentum
is $p=(E_d,0,0,-q_z/2)$ ($p'=(E_d,0,0,q_z/2)$), with
$q_z^2=Q^2$ and $E_d=\sqrt{Q^2/4+m_d^2}$.  In the zero-binding 
limit,\footnote{The difference between the proton and neutron masses
is neglected in addition to the deuteron binding energy,
the two being of the same order.}
$m_d=2m$ and the individual nucleon four-momenta are
$p_p=p_n=p/2$ and $p'_p=p'_n=p'/2$.  The hadronic
matrix elements are given by
\be
G_{\lambda'_d,\lambda_d}^\mu=\sum_{\lambda'_p,\lambda'_n}
    C_{\lambda'_p\lambda'_n}^{\lambda'_d} M^\mu(\lambda'_p,\lambda'_n,\lambda_d).
\ee
The initial spinors are as in (\ref{eq:initialspinors}); the final spinors are
specified by
\be
u'_N=\frac{\fsl{p}'/2+m}{\sqrt{E_d/2+m}}
  \left(\begin{array}{c} \phi^{(\lambda'_N)}(\hat z) \\ 0 \end{array}\right).
\ee
%

The three form factors are then extracted as~\cite{Arnold,SpinOne}
\be
G_C=\frac{-1}{2m_d\sqrt{1+\eta}}\frac{G_{00}^+-2G_{+-}^+}{3}, \;\;
G_M=\frac{2}{2m_d\sqrt{1+\eta}}\frac{G_{+0}^x}{\sqrt{2\eta}}, \;\;
G_Q=\frac{-1}{2m_d\sqrt{1+\eta}}\frac{G_{00}^++G_{+-}^+}{2\eta},
\ee
with $\eta\equiv \frac{Q^2}{4m_d^2}$ and the $+$ superscript denoting the
light-front sum of the 0 and $z$ components. For the helicity matrix
elements, the model yields the following $Q^2$ dependence:
\bea
G_{00}^+&=&0.5588{\cal N}m\mQ^9\left[1+129.1 \frac{m^2}{Q^2}+{\cal O}(\frac{m^4}{Q^4})\right], \\
G_{+-}^+&=&-69.85{\cal N}m\mQ^{11}\left[1+4.8 \frac{m^2}{Q^2}+{\cal O}(\frac{m^4}{Q^4})\right], \\
G_{+0}^x&=&8.851{\cal N}m\mQ^{10}\left[1+4.8 \frac{m^2}{Q^2}+{\cal O}(\frac{m^4}{Q^4})\right],
\eea
with ${\cal N}$ the unknown normalization.  The factor of $m/Q$ associated 
with each helicity flip \cite{CarlsonGross} is clearly evident.
For the form factors, we find
\bea
G_C&=&-\frac{0.5588}{\sqrt{1+\eta}}\frac{{\cal N}}{12}\mQ^9\left[1+379.1\frac{m^2}{Q^2}+{\cal O}(\frac{m^4}{Q^4})\right], \\
G_M&=&\frac{8.851}{\sqrt{\eta(1+\eta)}}\frac{{\cal N}}{2\sqrt{2}}\mQ^{10}\left[1+4.8\frac{m^2}{Q^2}+{\cal O}(\frac{m^4}{Q^4})\right], \\
G_Q&=&-\frac{0.5588}{\eta\sqrt{1+\eta}}\frac{{\cal N}}{8}\mQ^9\left[1+4.086\frac{m^2}{Q^2}+{\cal O}(\frac{m^4}{Q^4})\right].
\eea
The leading $\pm$ signs are as expected for large $Q^2$.

We have left the kinematic factor $\eta=Q^2/16m^2$ without substitution, because there can be three 
regimes for $Q^2$.  In addition to $Q^2$ large or small, there can be an intermediate region
where $Q^2$ is large but $\eta$ is not.  Such an intermediate
regime does exist for $G_M$ and $G_Q$, where the coefficients of
the nonleading terms are small enough for this correction to
be small while $\eta$ is also small.  For $G_C$,  this is
not the case, because the coefficient of the nonleading term is
large enough to require a $Q^2$ value for which $\eta$ is also large.
In the intermediate regime, we obtain
\be
G_M\sim\mQ^{11}, \;\; G_Q\sim\mQ^{11},
\ee
and for the large-$\eta$ regime
\be
G_C\sim\mQ^{10},\;\;G_M\sim\mQ^{12}, \;\; G_Q\sim\mQ^{12}.
\ee

\begin{figure}[ht]
\vspace{0.2in}
\centerline{\includegraphics[width=10cm]{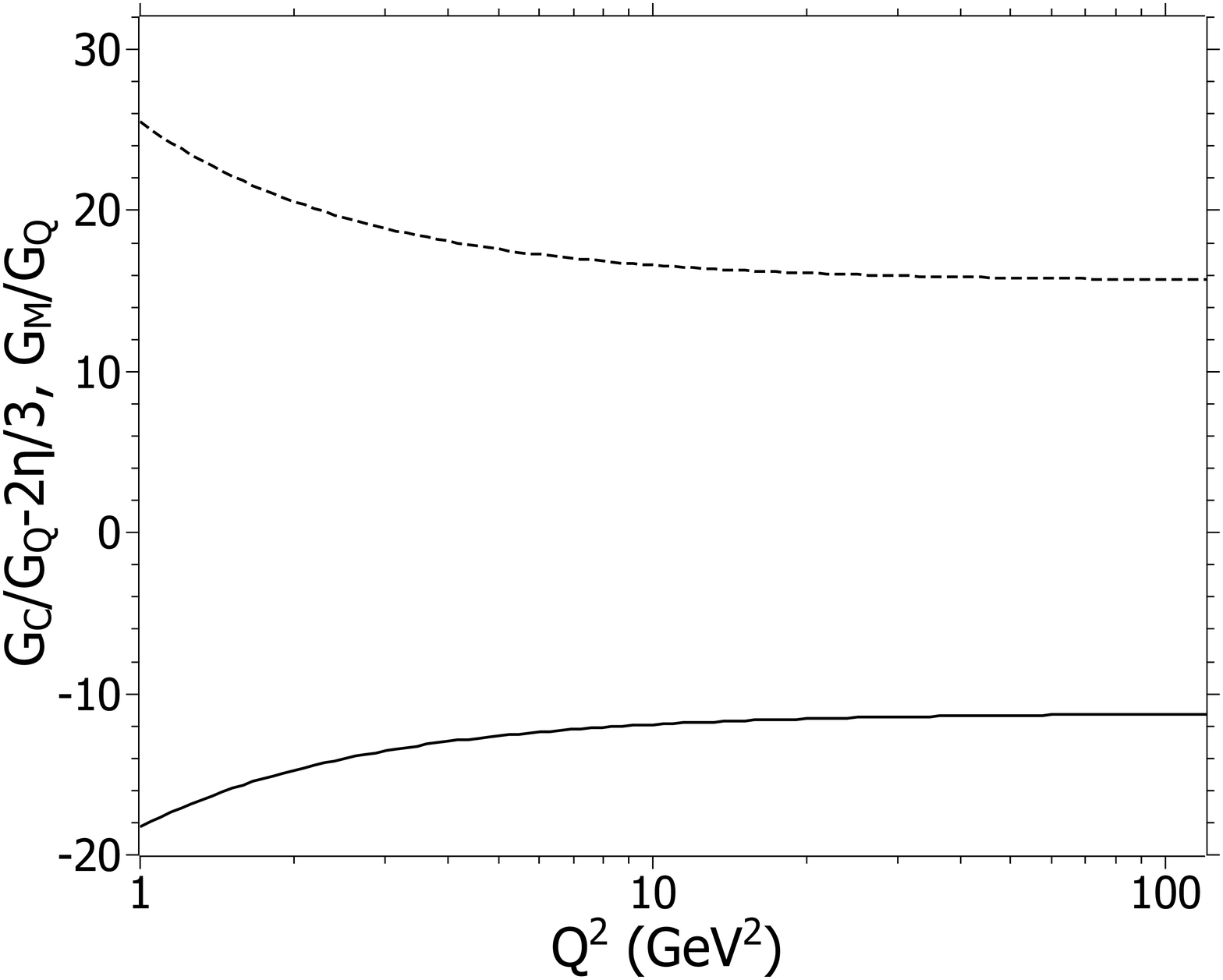}}
\caption{\label{fig:FFratios} Ratios $G_C/G_Q-2\eta/3$ (dashed) and $G_M/G_Q$ (solid) for the model deuteron form factors.
}
\end{figure}

Ratios of these form factors at very large $Q^2$ can be 
compared with the tree-level ratios for a point-like
spin-one particle, such as the $W^+$, which are~\cite{SpinOne}
\be
\frac{G_C}{G_Q}=\frac23\eta-1,\;\;
\frac{G_M}{G_Q}=-2.
\ee
Such behavior is immediately reproduced for form factors separated according
to a Drell--Yan frame~\cite{DrellYan}, with the assumption of strict
$G^+_{00}$ dominance~\cite{SpinOne}.
In terms of our hadronic matrix elements, we have
\be
\frac{G_C}{G_Q}=\frac23\eta-2\eta\frac{G^+_{+-}}{G^+_{00}+G^+_{+-}}, \;\;
\frac{G_M}{G_Q}=-2\sqrt{2\eta}\frac{G^x_{+0}}{G^+_{00}+G^+_{+-}}.
\ee
As already observed in \cite{SpinOne}, these Breit-frame ratios cannot
both be resolved by simply assuming $G^+_{00}$ dominance.  From our
model, we obtain
\be
\frac{G_C}{G_Q}=\frac23\eta+15.6+{\cal O}(\frac{m^2}{Q^2}),\;\;
\frac{G_M}{G_Q}=-11.2+{\cal O}(\frac{m^2}{Q^2}).
\ee
The leading $\frac23\eta$ is just kinematic.  The deviations of
15.6 and -11.2 from -1 and -2, respectively, are due to nonleading
contributions multiplied by powers of $\eta$.  Similar deviations
will arise for calculations done in the Drell--Yan frame, because
$\eta$ factors again interfere with strict $G_{00}^+$ dominance.
Plots of these ratios are shown in Fig.~\ref{fig:FFratios}.

\subsection{Structure functions} \label{sec:structurefunctions}

\begin{figure}[ht]
\vspace{0.2in}
\centerline{\includegraphics[width=10cm]{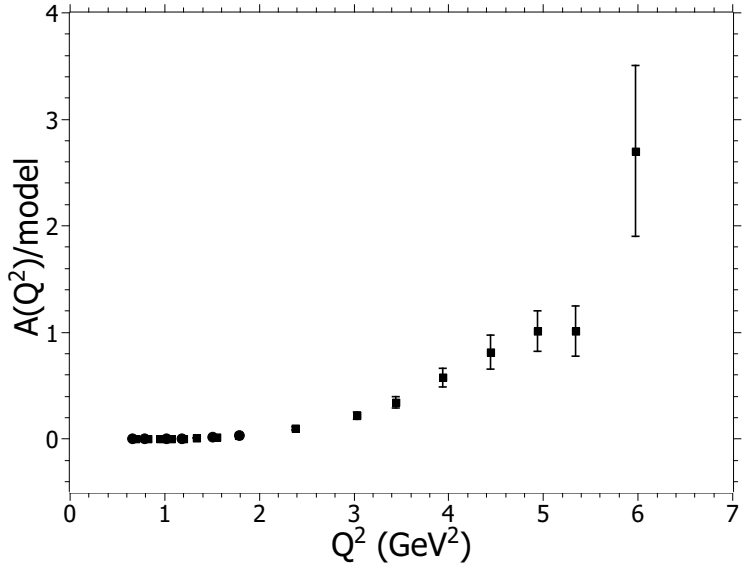}}
\caption{\label{fig:Adata/model} Data for the deuteron structure function
$A(Q^2)$ divided by the model function, including an arbitrary normalization.
Experimental values are taken from \protect\cite{AbbotA} (circles) and 
\protect\cite{Alexa} (squares).
}
\end{figure}

\begin{figure}[ht]
\vspace{0.2in}
\centerline{\includegraphics[width=10cm]{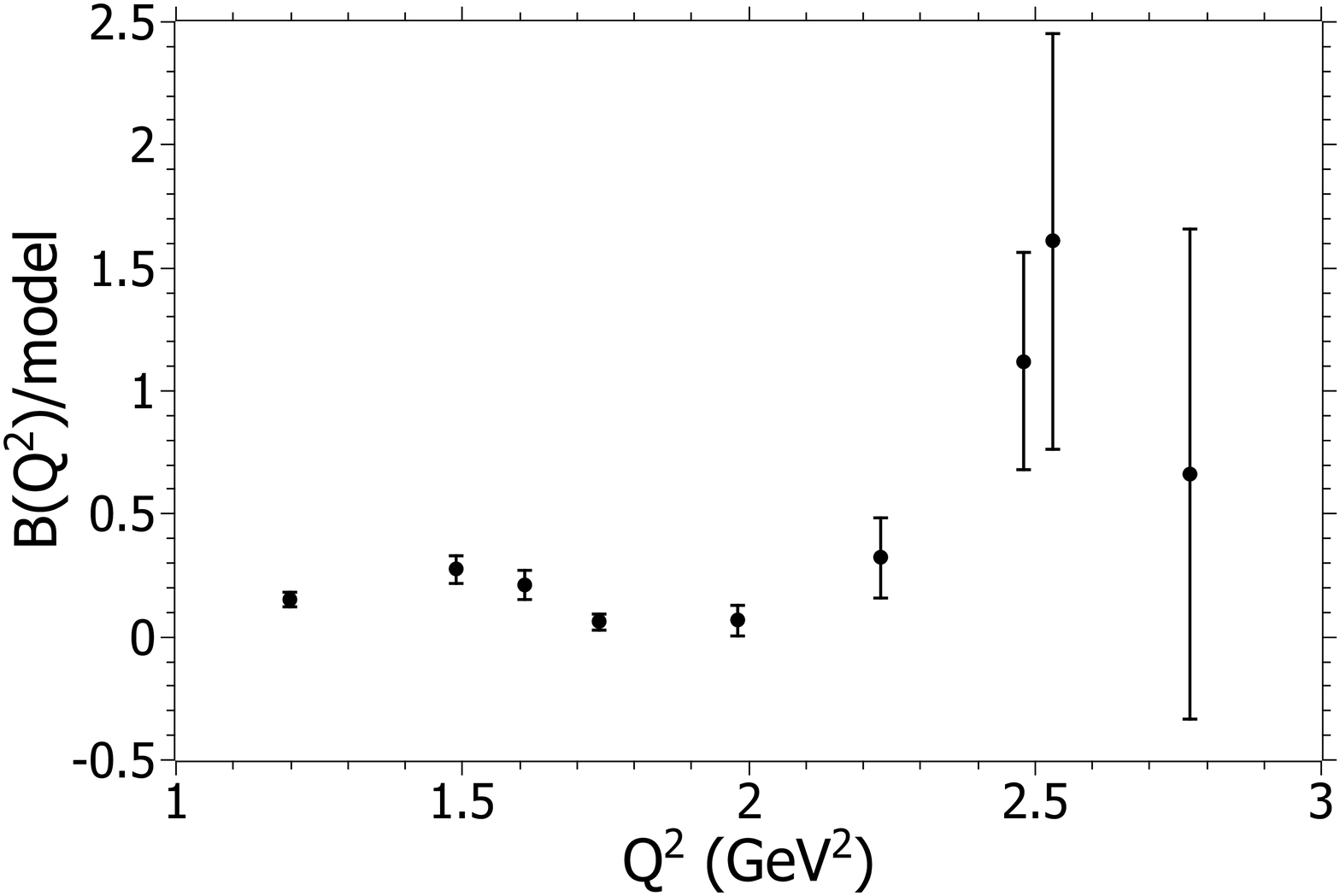}}
\caption{\label{fig:Bdata/model} Data for the deuteron structure function
$B(Q^2)$ divided by the model function, including an arbitrary normalization.
Experimental values are taken from \protect\cite{Bosted}.
}
\end{figure}

Experiments designed to extract these form factors measure cross sections
and polarization observables in elastic electron-deuteron scattering.
The unpolarized cross section 
\be
\frac{d\sigma}{d\Omega}\propto S,\;\; S\equiv A(Q^2)+B(Q^2)\tan^2(\theta_e/2)
\ee
depends on the electron scattering angle $\theta_e$ and two structure functions
\bea
A(Q^2)&\equiv&G_C^2+\frac89\eta^2 G_Q^2+\frac23\eta G_M^2, \\
B(Q^2)&\equiv& \frac43\eta(1+\eta)G_M^2.
\eea
These have been measured at the highest $Q^2$ yet attained at JLab~\cite{Bosted,AbbotA,Alexa},
and $A$ has been measured at comparable $Q^2$ at SLAC~\cite{Arnoldetal}.  However,
these do not yet reach the $Q^2$ values needed for a definitive comparison.
Figures~\ref{fig:Adata/model} and \ref{fig:Bdata/model} show plots of the data
divided by the model, including an arbitrary normalization factor.

In our model, expansions of these functions in inverse powers of $Q^2$ are
\bea
A(Q^2)&=&0.1041{\cal N}^2\mQ^{20}\left[1+1246\frac{m^2}{Q^2}+{\cal O}(\frac{m^4}{Q^4})\right], \\
B(Q^2)&=&13.06{\cal N}^2\mQ^{20}\left[1+9.6\frac{m^2}{Q^2}+{\cal O}(\frac{m^4}{Q^4})\right].
\eea
Because the expansion for $G_C$ is valid only for large $\eta$, we
have used the explicit form of $\eta$ in constructing the expansion for $A$.
The function $B$ is independent of $\eta$; the leading factor of $\eta(1+\eta)$ in
its definition exactly cancels against factors in the relationship of $G_M$ to
hadronic matrix elements.  The expansion for $B$ converges much faster than
the expansion for $A$, and the leading $Q^2$ behavior is dominant for $Q^2\gg 10$ GeV$^2$
only for $B$.  For $A$, one must wait until impossibly large $Q^2$, which enters
a regime where the collective quark substructure is important, including 
hidden-color effects~\cite{BrodskyJiLepage}, and the point-like approximation
used in our model is invalid. 

\begin{figure}[ht]
\vspace{0.2in}
\centerline{\includegraphics[width=10cm]{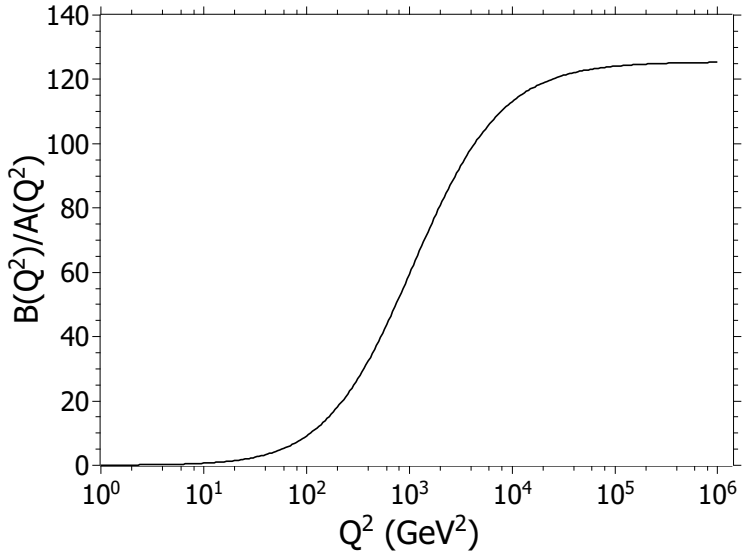}}
\caption{\label{fig:BAratio} Ratio of $B$ to $A$ for the model deuteron structure functions.
}
\end{figure}

In \cite{Bosted} the large $Q^2$ behavior of $B$ is quoted as being $Q^{-24}$
from perturbative QCD.  This faster fall off compared to $A$ is attributed to
the extra suppression of the helicity flip involved in $G_M$.  However, there
are other compensating factors, and, just as in our model, the behavior of
$B$ should be $Q^{-20}$, which is the same as $A$.  In Fig.~\ref{fig:BAratio}
we plot the ratio of $B$ to $A$ for a large range of $Q^2$.  This ratio
becomes constant at very large $Q^2$.  Although the
plots begin at low $Q^2$, there is nothing in the model that could reproduce
diffractive minima, hence the smooth appearance.

\subsection{Tensor polarization observables} \label{sec:polarization}

\begin{figure}[ht]
\vspace{0.2in}
\centerline{\includegraphics[width=10cm]{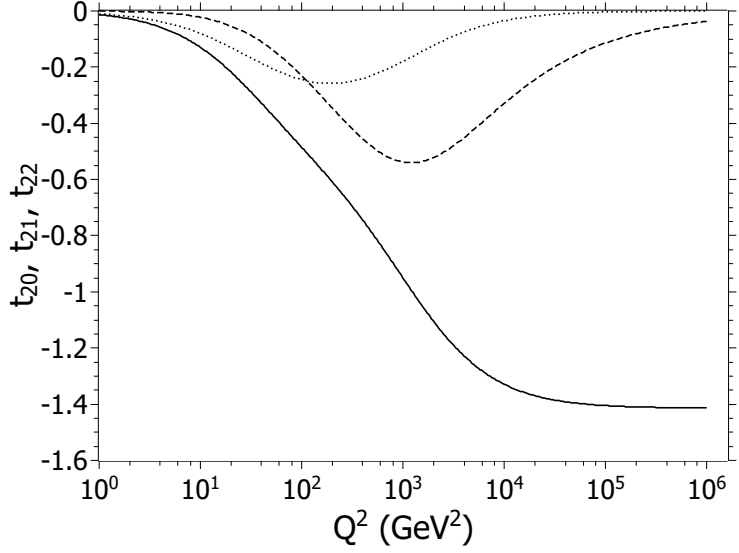}}
\caption{\label{fig:txx0} Deuteron tensor polarization observables
$t_{20}$ (solid), $t_{21}$ (dashed), and $t_{22}$ (dotted) as computed in the model at an
angle of $\theta_e=0^\circ$.  The asymptotic value of $t_{20}(0^\circ)$
is $-\sqrt{2}$, as predicted by perturbative QCD~\protect\cite{Dymarz,CarlsonGross}.
}
\end{figure}

\begin{figure}[ht]
\vspace{0.2in}
\centerline{\includegraphics[width=10cm]{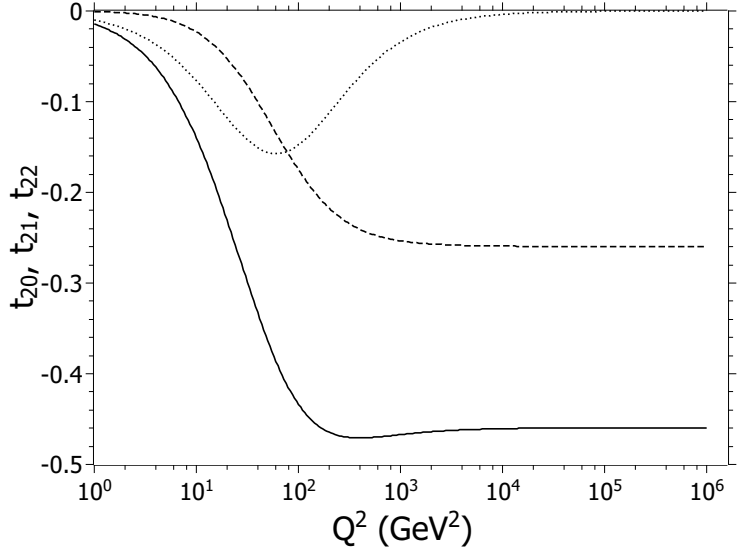}}
\caption{\label{fig:txx30} Same as Fig.~\protect\ref{fig:txx0} but
for an angle of $\theta_e=30^\circ$.
}
\end{figure}

\begin{figure}[ht]
\vspace{0.2in}
\centerline{\includegraphics[width=10cm]{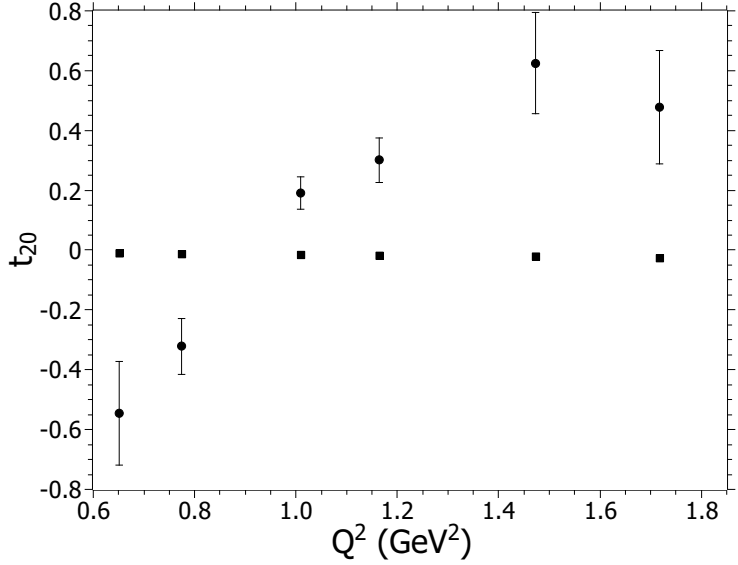}}
\caption{\label{fig:t20data} Plots of the tensor polarization observable $t_{20}$
of the deuteron from both data~\protect\cite{AbbotT} (circles) and the
model (squares) considered in the text. The angle $\theta_e$ varies and is as follows
in order of increasing $Q^2$: 35.6$^\circ$, 33.4$^\circ$, 29.8$^\circ$,
27.3$^\circ$, 23.0$^\circ$, and 19.8$^\circ$.
}
\end{figure}

\begin{figure}[ht]
\vspace{0.2in}
\centerline{\includegraphics[width=10cm]{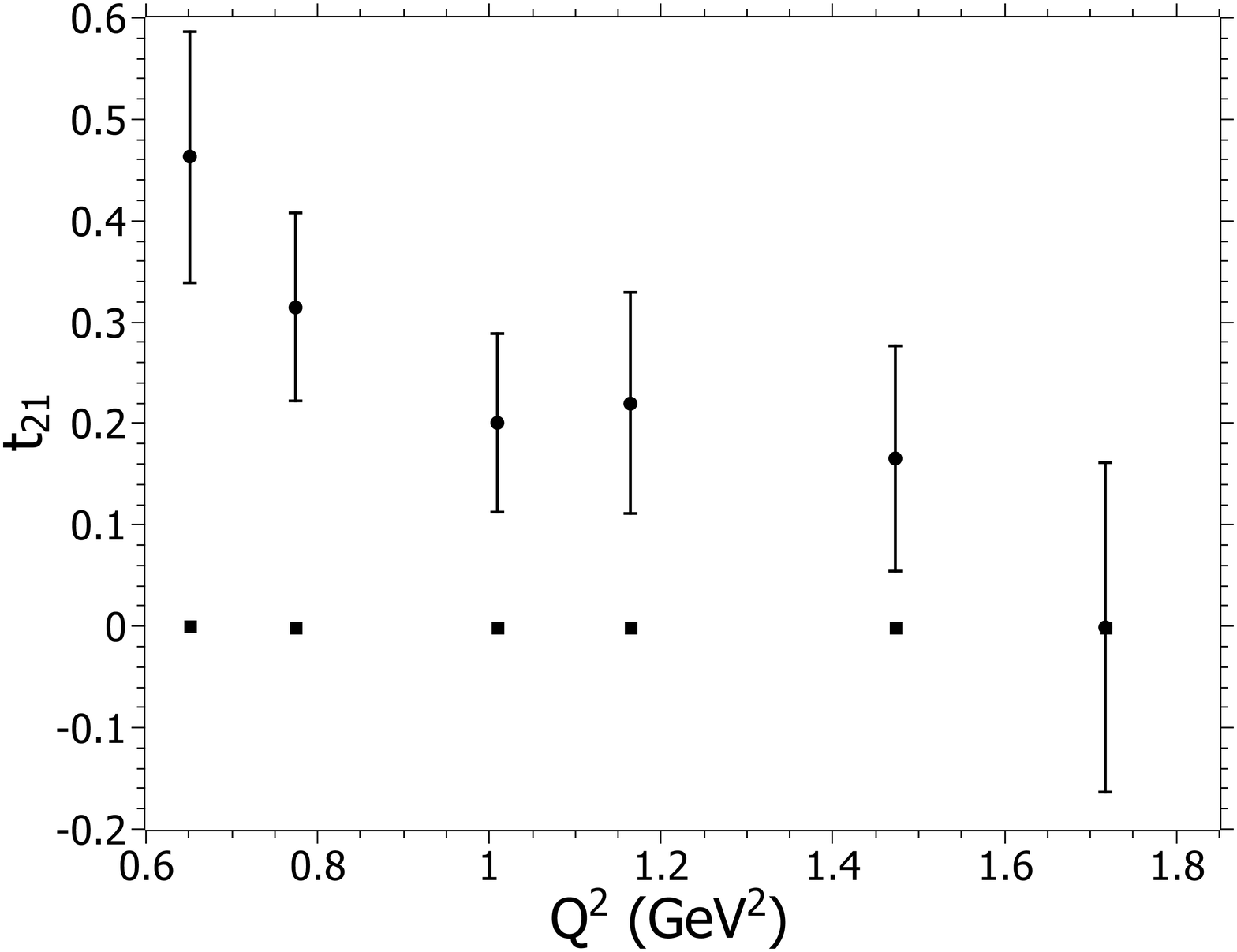}}
\caption{\label{fig:t21data} Same as Fig.~\protect\ref{fig:t20data} but
for $t_{21}$.
}
\end{figure}

\begin{figure}[ht]
\vspace{0.2in}
\centerline{\includegraphics[width=10cm]{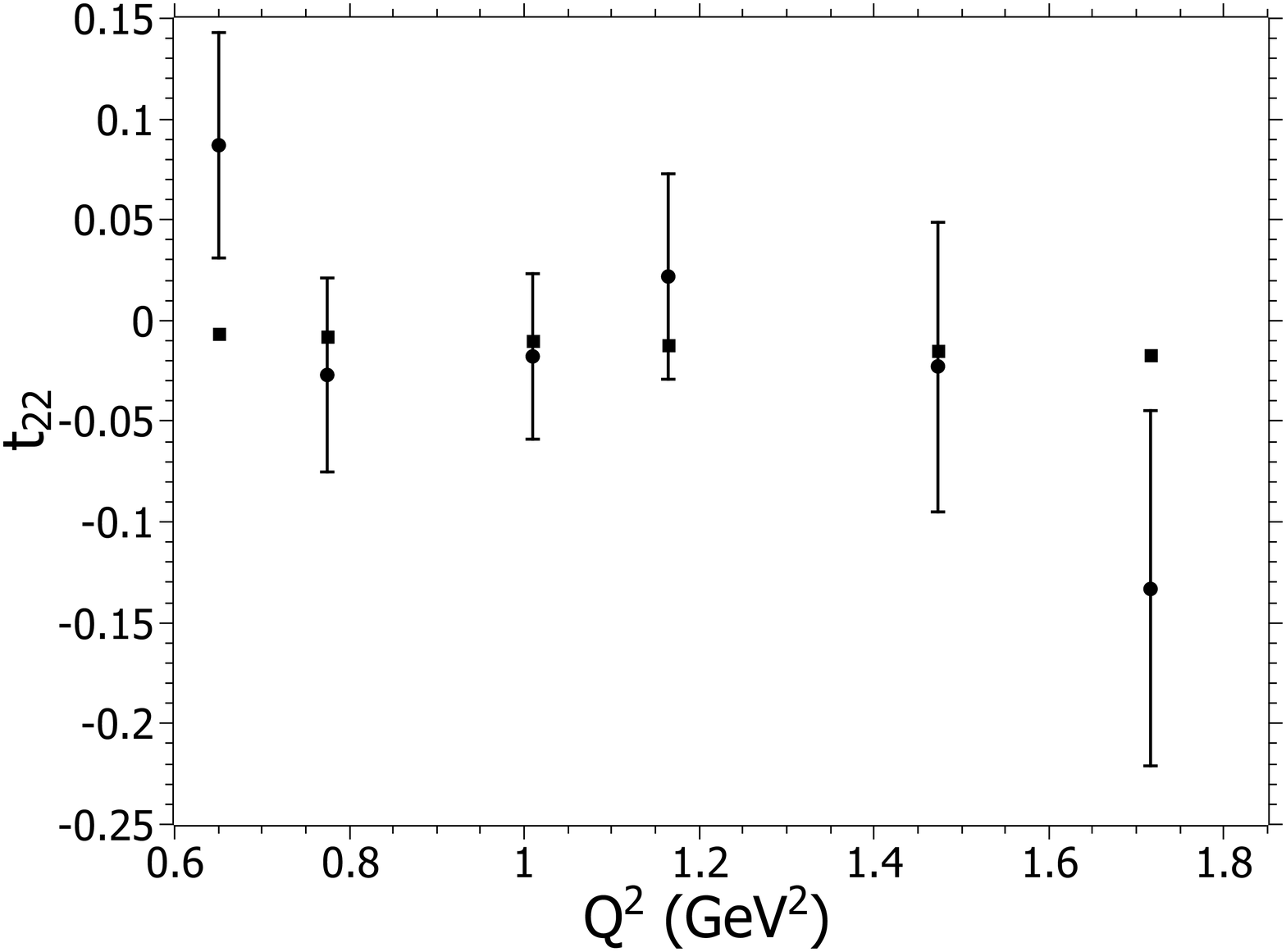}}
\caption{\label{fig:t22data} Same as Fig.~\protect\ref{fig:t20data} but
for $t_{22}$.
}
\end{figure}

Experiments can also extract tensor polarization observables~\cite{GilmanGross,AbbotT} 
\bea
t_{20}&\equiv&-\frac{1}{\sqrt{2}S}
\left[\frac83\eta G_C G_Q+\frac89\eta^2 G_Q^2+\frac13\eta\left(1+2(1+\eta)\tan^2(\theta_e/2)\right)G_M^2\right], \\
t_{21}&\equiv&\frac{2\eta}{\sqrt{3}S\cos(\theta_e/2)}\sqrt{\eta+\eta^2\sin^2(\theta_e/2)}G_M G_Q, \\
t_{22}&\equiv&-\frac{\eta}{2\sqrt{3}S}G_M^2.
\eea
The highest $Q^2$ measurements of these were also done at JLab~\cite{AbbotT}.
When $\eta$ is held explicit, expansions in $m/Q$ are
\bea
t_{20}&=&-\sqrt{2}+1064[1+2(1+\eta)\tan^2(\theta_e/2)]\mQ^2+{\cal O}(\frac{m^4}{Q^4}), \\
t_{21}&=&38.8\sec(\theta_e/2)\sqrt{\eta+\sin^2(\theta_e/2)}\frac{m}{Q} \\
      && +\sec(\theta_e/2)\sqrt{\eta+\sin^2(\theta_e/2)}[48606+77869(1+\eta)\tan^2(\theta_e/2)]\mQ^3+{\cal O}(\frac{m^4}{Q^4}),
      \nonumber \\
t_{22}&=&-434.5\mQ^2+[544703+872133(1+\eta)\tan^2(\theta_e/2)]\mQ^4+{\cal O}(\frac{m^6}{Q^6}).
\eea
While at very large $Q^2$, they are
\bea
t_{20}&=&-\sqrt{2}+133\tan^2(\theta_e/2)+1064[1+2\tan^2(\theta_e/2)]\mQ^2+{\cal O}(\frac{m^4}{Q^4}), \\
t_{21}&=&1217 \sec(\theta_e/2)\sin(\theta_e/2)[\tan^2(\theta_e/2)-0.007972]
      +\left[\right]\mQ^2+{\cal O}(\frac{m^4}{Q^4}), \\
t_{22}&=&-[434.5-54508\tan^2(\theta_e/2)]\mQ^2 \\
     &&+[544703+872133\tan^2(\theta_e/2)]\mQ^4+{\cal O}(\frac{m^6}{Q^6}). \nonumber
\eea
The coefficients of nonleading terms are quite large.  Thus, very
large $Q^2$ is required for the leading term to be dominant,
well beyond any available data.
The limit of $-\sqrt{2}$ for $t_{20}$ at $\theta_e=0^\circ$ was
an early prediction of perturbative QCD~\cite{Dymarz,CarlsonGross}.
However, as argued elsewhere~\cite{SpinOne}, this value is
obtained only at very large $Q^2$, and the value is quite
different for small $\eta$.
Figures~\ref{fig:txx0} and \ref{fig:txx30} show plots of these 
observables at angles of 0$^\circ$ and 30$^\circ$, respectively.
We also compare with data~\cite{AbbotT} in Figs.~\ref{fig:t20data}, \ref{fig:t21data},
and \ref{fig:t22data}.  At these `small'  values of $Q^2$, only
$t_{22}$ is consistent with data, something which is likely accidental
with both data and model values near zero.

\section{Photodisintegration}
\label{sec:photodis}

\begin{figure}[ht]
\vspace{0.2in}
\centerline{\includegraphics[width=10cm]{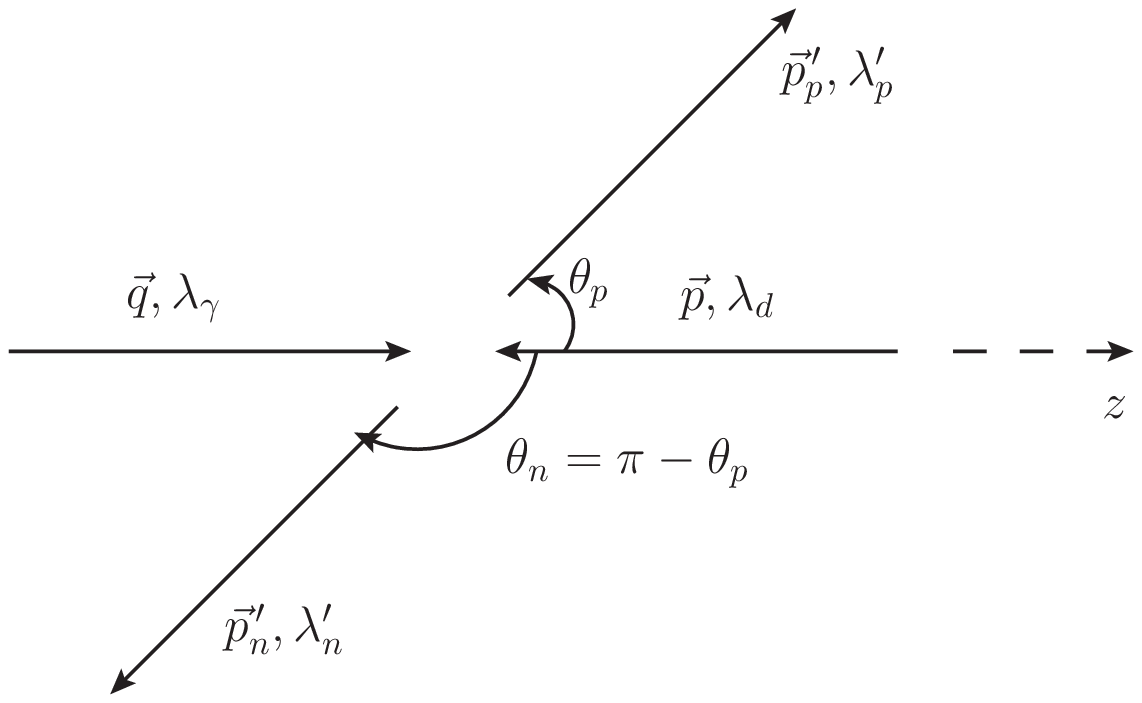}}
\caption{\label{fig:photodis} Kinematics for deuteron photodisintegration
in the c.m.\ frame, with $\vec{q}$ the photon momentum and $\vec{p}=-\vec{q}$
the deuteron momentum.  The final proton and neutron momenta are $\vec{p}^{\,\prime}_p$
and $\vec{p}^{\,\prime}_n$.  The $\lambda$'s are helicities. Coordinates are 
chosen such that the photon enters along the positive $z$ direction and
the azimuthal angle $\phi_p$ of the proton is zero.
}
\end{figure}

In the photodisintegration of a deuteron, a real photon
is absorbed and the two constituent nucleons emitted.
This process is depicted in Fig.~\ref{fig:photodis}.
The initial deuteron and photon four-momenta in the
center-of-mass (c.m.) frame are
$p = (E_d, 0, 0, -q_z)$ and $q = (q_z, 0, 0, q_z)$,
where the incident photon is taken along the positive
$z$ axis.  The final proton and neutron four-momenta
are $p'_p= (E'_p, \vec p^{\,\prime}_p)$ and $p'_n= (E'_n, \vec p^{\,\prime}_n)$,
with $\theta_p$ and $\phi_p$ the polar and azimuthal 
angles of the final proton.  By ignoring the nucleon
mass difference, we have $E'_p=E'_n$, because momentum conservation
guarantees $\vec p^{\,\prime}_n=-\vec p^{\,\prime}_p$ in the c.m. frame.

In terms of the Mandelstam variable $s$, the c.m.\ energies
and momenta are
\be
E_d = (s + 4 m^2)/(2\sqrt{s}), \;\;
q_z = (s - 4 m^2)/(2 \sqrt{s}), \;\;
E'_p = \sqrt{s}/2, \;\;
|\vec p^{\,\prime}_p| = \sqrt{s - 4 m^2}/2.
\ee
The photon energy in the lab frame is $E_\gamma=(s-4m^2)/4m$.
We will work at large $s$, so that momentum transfers are large.  

The standard definition of helicity amplitudes for
photodisintegration is~\cite{Barannik}
\be
F_{i\pm} \equiv \epsilon_\nu(\lambda_\gamma) M^\nu(\lambda'_p,\lambda'_n,\lambda_d)
\ee
with $\epsilon$ the polarization vector for a photon with 
helicity $\lambda_\gamma$ and $M^\nu$ given in (\ref{eq:Mnu}).  The index
$i$ is associated with particular helicity combinations as follows:
\bea
F_{1\pm}&=&\epsilon_\nu(1) M^\nu(\pm\frac12,\pm\frac12,1), \;\;
F_{2\pm}=\epsilon_\nu(1) M^\nu(\pm\frac12,\pm\frac12,0), \\
F_{3\pm}&=&\epsilon_\nu(1) M^\nu(\pm\frac12,\pm\frac12,-1), \;\;
F_{4\pm}=\epsilon_\nu(1) M^\nu(\pm\frac12,\mp\frac12,1), \\
F_{5\pm}&=&\epsilon_\nu(1) M^\nu(\pm\frac12,\mp\frac12,0), \;\;
F_{6\pm}=\epsilon_\nu(1) M^\nu(\pm\frac12,\mp\frac12,-1).
\eea
The other helicity combinations are related to these
by parity.

The helicity amplitudes can be used to compute various
polarization observables.  The recoil-proton polarization $P_y$
measures the asymmetry parallel/antiparallel to the 
normal $\hat y \propto \vec q \times \vec p^{\,\prime}_p$ to the scattering plane:
\be
P_y=2 {\rm Im}\sum_{i=1}^3[F_{i+}^\dagger F_{i+3,-}+F_{i+3,+}^\dagger F_{i-}]/f(\theta),
\ee
where $f(\theta)=\sum_{i=1}^6[|F_{i+}|^2+|F_{i-}|^2]$ is the sum of all the
helicity amplitudes squared.
The transferred polarizations $C_{x'}$ and $C_{z'}$ measure asymmetries 
parallel/antiparallel to the $\hat x'\propto\vec p^{\,\prime}_p\times\hat y$ and 
$\hat z'=\hat p^{\,\prime}_p$ directions:
\bea
C_{x'}&=&2{\rm Re}\sum_{i=1}^3[F_{i+}^\dagger F_{i+3,-}+F_{i+3,+}^\dagger F_{i-}]/f(\theta), \\
C_{z'}&=&\sum_{i=1}^6[|F_{i+}|^2-|F_{i-}|^2]/f(\theta).
\eea
The asymmetry $\Sigma$ for linearly polarized photons is given by
\be
\Sigma=-2{\rm Re}\left[\sum_{\pm}(F_{1\pm}^\dagger F_{3\mp}-F_{4\pm}^\dagger F_{6\mp})
-F_{2+}^\dagger F_{2-}+F_{5+}^\dagger F_{5-}\right]/f(\theta).
\ee
Each observable is formed as a ratio, which sets aside questions of normalization.

Because we only need to consider photons with helicity +1, the polarization 
vector is always $\epsilon=-\frac{1}{\sqrt{2}}(0,1,i,0)$, relative to the
momentum in the positive $z$ direction.  The final Dirac spinors are
\be
u'_N=\frac{\fsl{p}'_N+m}{\sqrt{E'_N+m}}
  \left(\begin{array}{c} \phi^{(\lambda'_N)}(\hat p'_N) \\ 0 \end{array}\right),
\ee
with $\theta_n=\pi-\theta_p$, $\phi_n=\phi_p+\pi=\pi$, and
\be
\phi^{(1/2)}(\hat p'_N)=
   \left(\begin{array}{c} \cos(\theta_N/2) \\ e^{i\phi_N}\sin(\theta_N/2) \end{array}\right), \;\;
\phi^{(-1/2)}(\hat p'_N)=
   \left(\begin{array}{c} -e^{-i\phi_N}\sin(\theta_N/2) \\ \cos(\theta_N/2) \end{array}\right).
\ee

\begin{figure}[ht]
\vspace{0.2in}
\begin{center}
\begin{tabular}{cc}
\includegraphics[width=8cm]{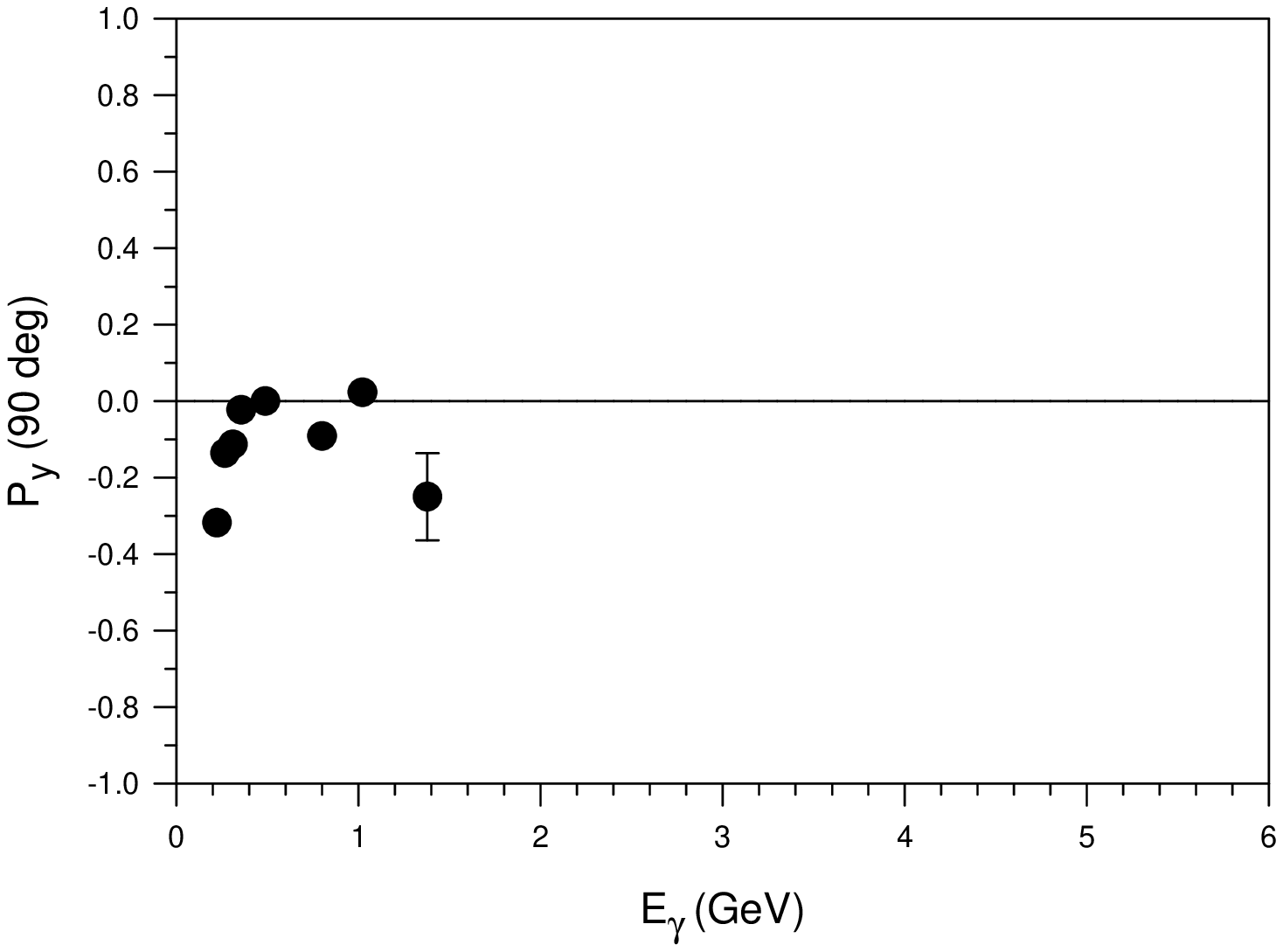} &
\includegraphics[width=8cm]{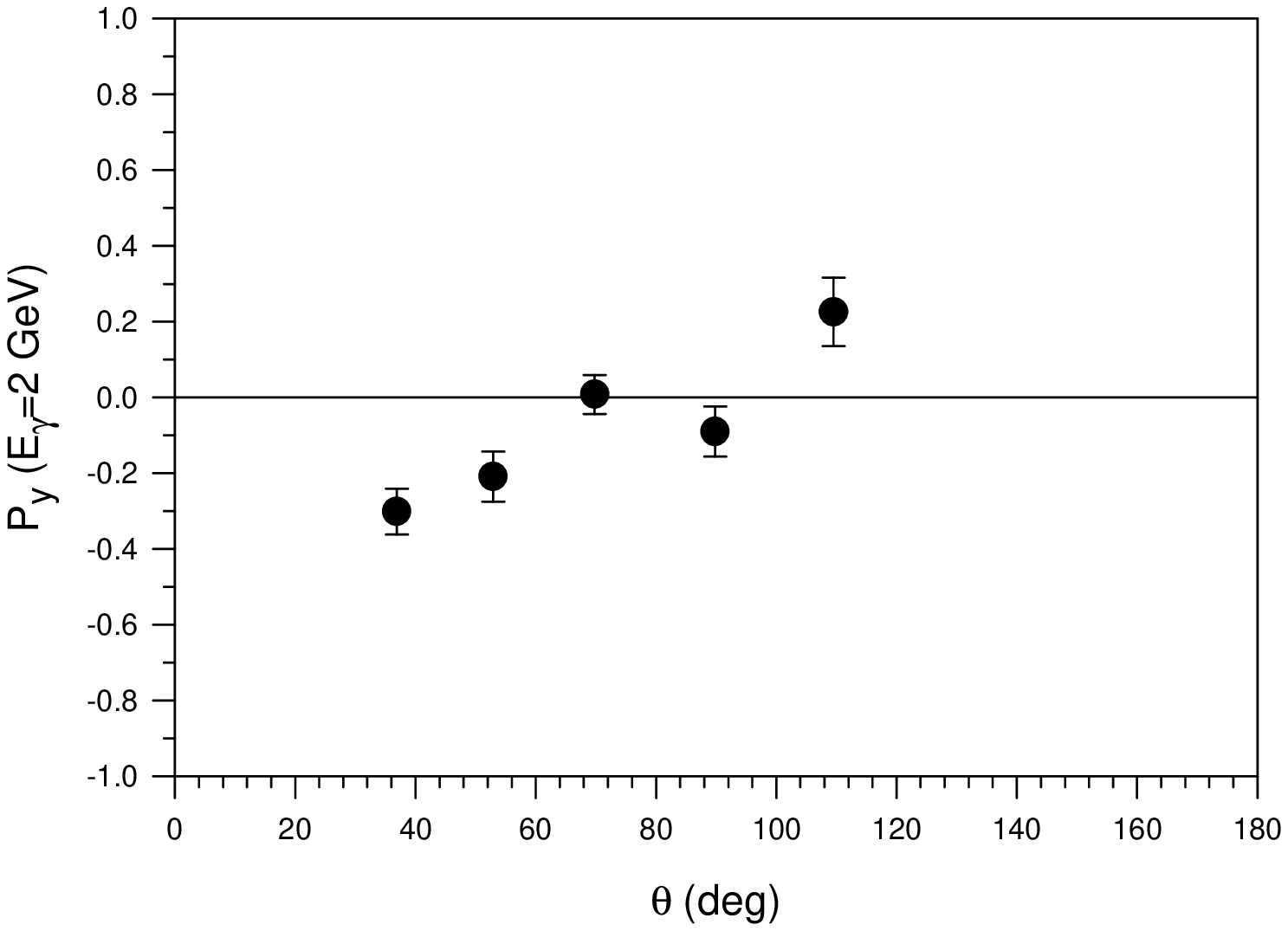} \\
(a) & (b)
\end{tabular}
\end{center}
\caption{\label{fig:Py} Recoil proton polarization $P_y$ as a function 
of (a) photon energy $E_\gamma$ and (b) proton angle $\theta$.  For the latter,
the photon energy is 2 GeV.  The solid line is the RNHA prediction;
the data points are from \protect\cite{EgammaData,AngleData}.
}
\end{figure}

\begin{figure}[ht]
\vspace{0.2in}
\begin{center}
\begin{tabular}{cc}
\includegraphics[width=8cm]{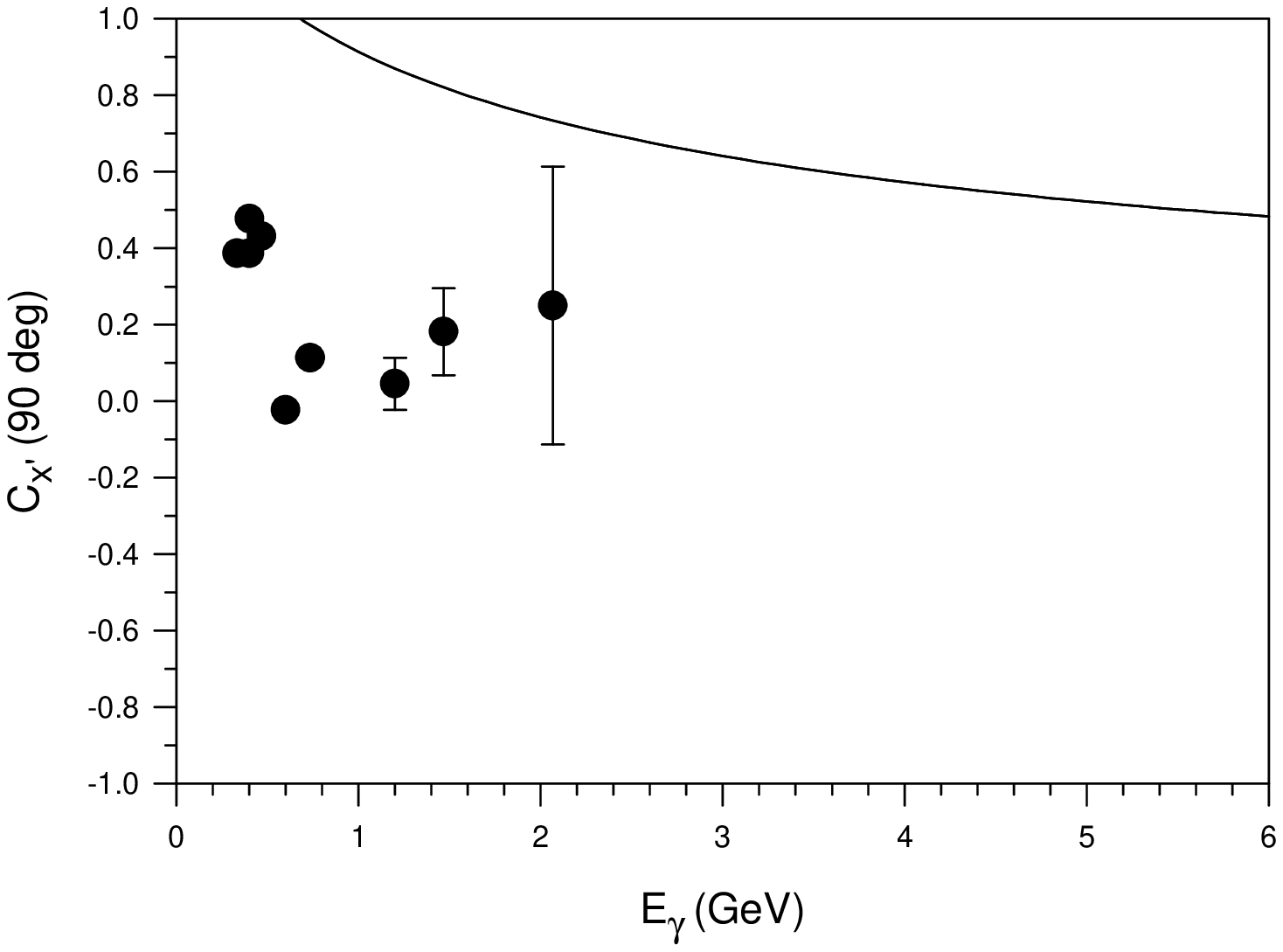} &
\includegraphics[width=8cm]{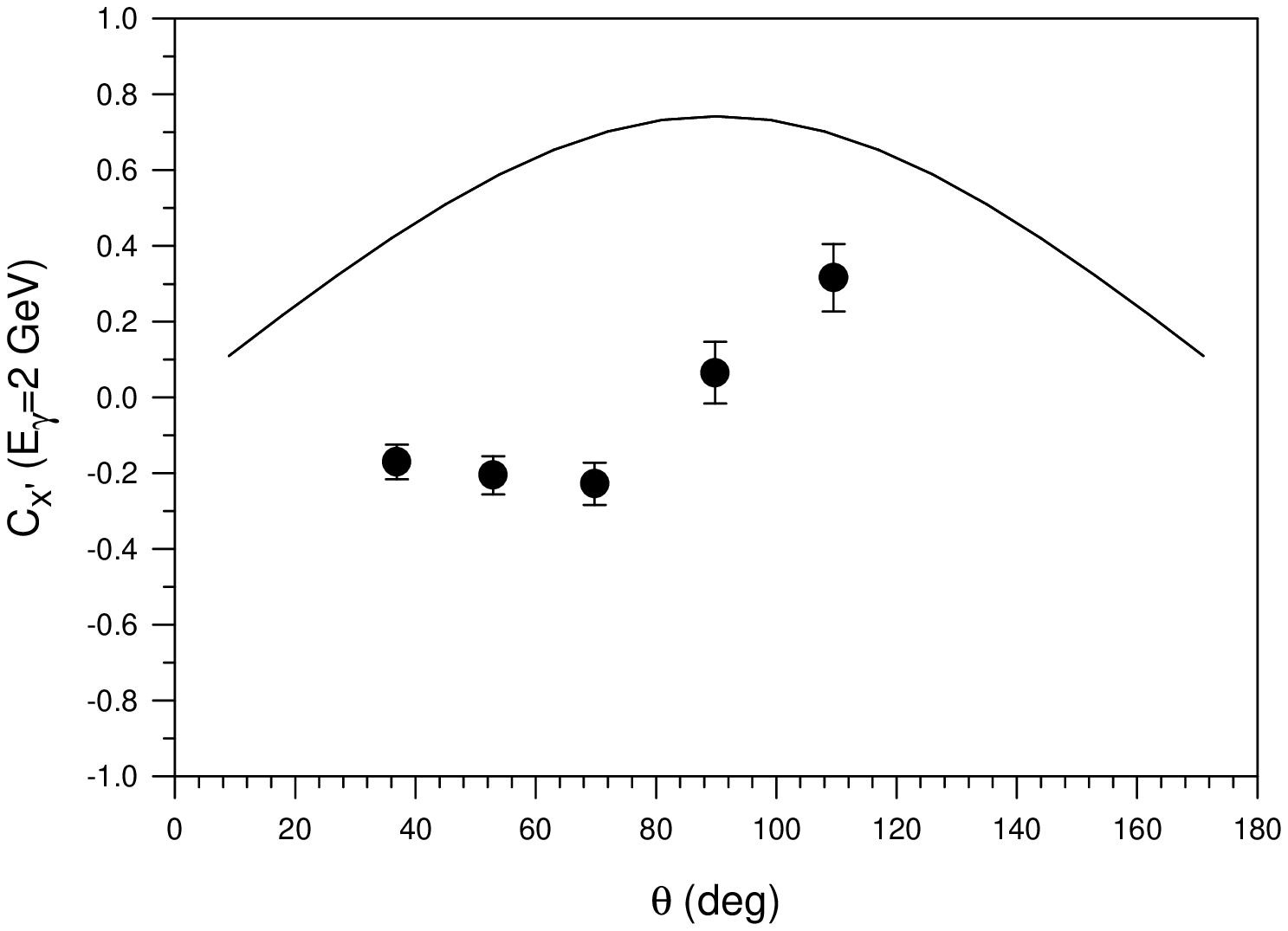} \\
(a) & (b)
\end{tabular}
\end{center}
\caption{\label{fig:Cxprime} Same as Fig.~\ref{fig:Py} but for 
the transferred polarization $C_{x'}$.
}
\end{figure}

\begin{figure}[ht]
\vspace{0.2in}
\begin{center}
\begin{tabular}{cc}
\includegraphics[width=8cm]{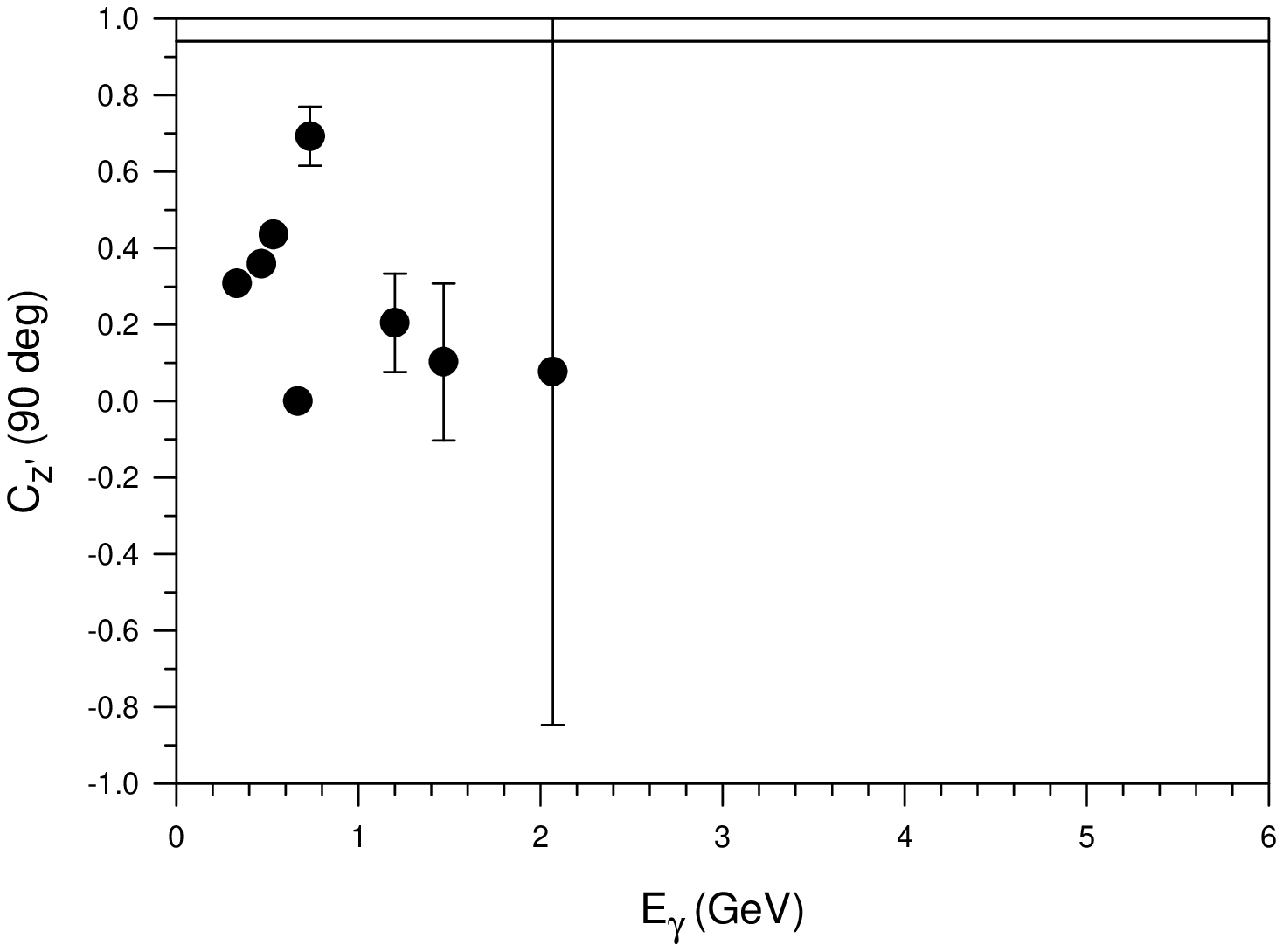} &
\includegraphics[width=8cm]{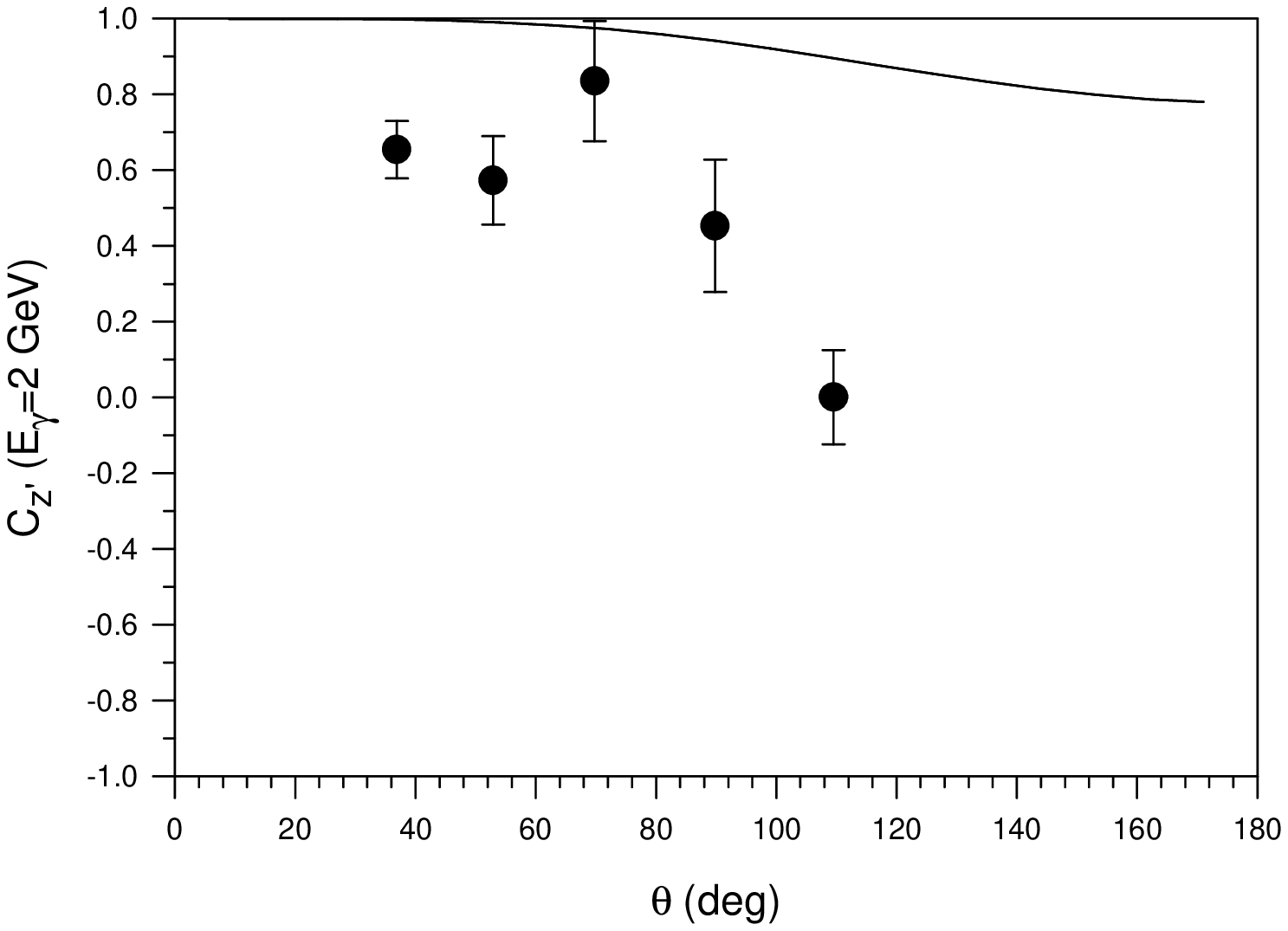} \\
(a) & (b)
\end{tabular}
\end{center}
\caption{\label{fig:Czprime} Same as Fig.~\ref{fig:Cxprime}
but for $C_{z'}$.
}
\end{figure}

\begin{figure}[ht]
\vspace{0.2in}
\begin{center}
\begin{tabular}{cc}
\includegraphics[width=8cm]{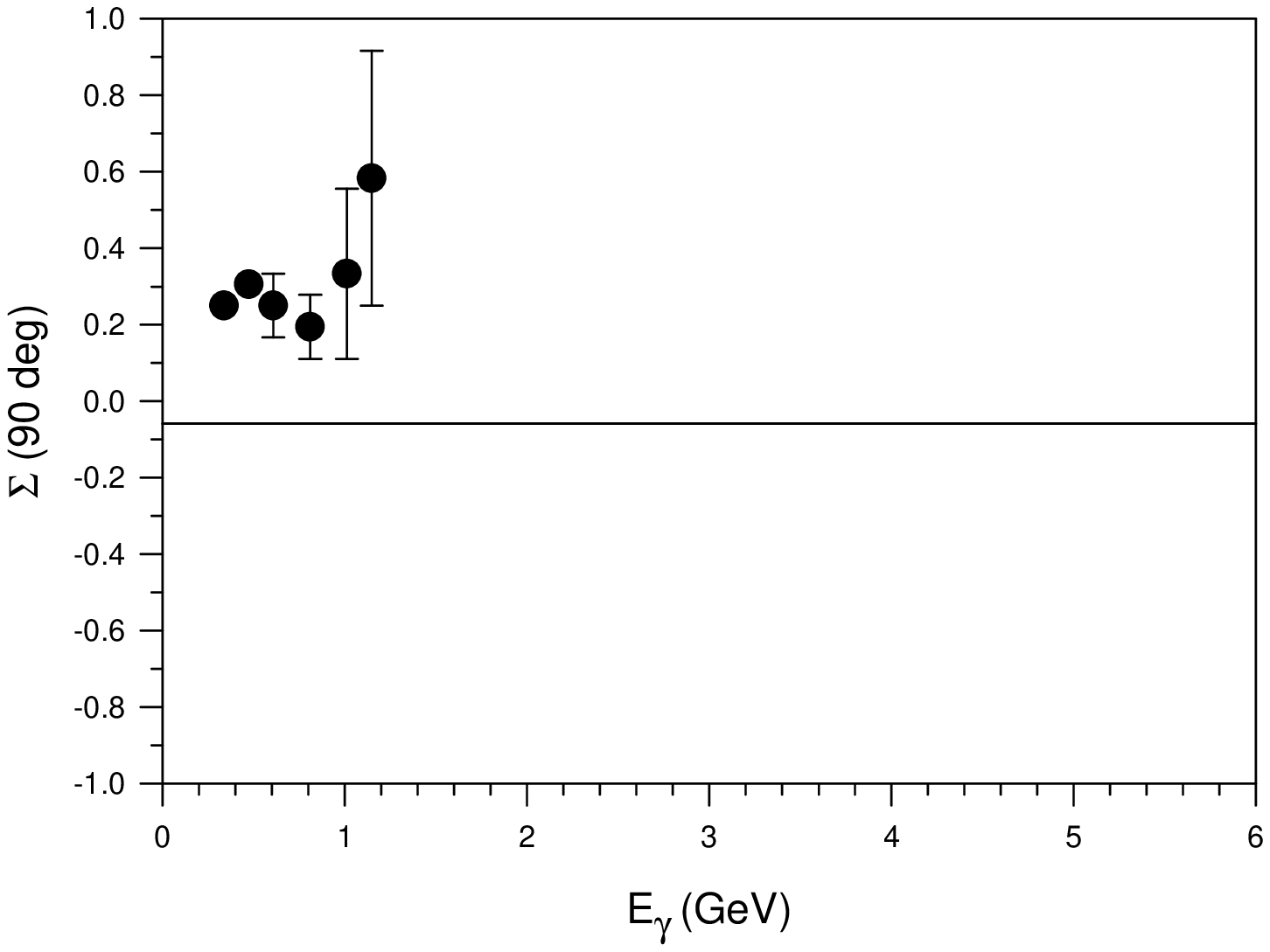} &
\includegraphics[width=8cm]{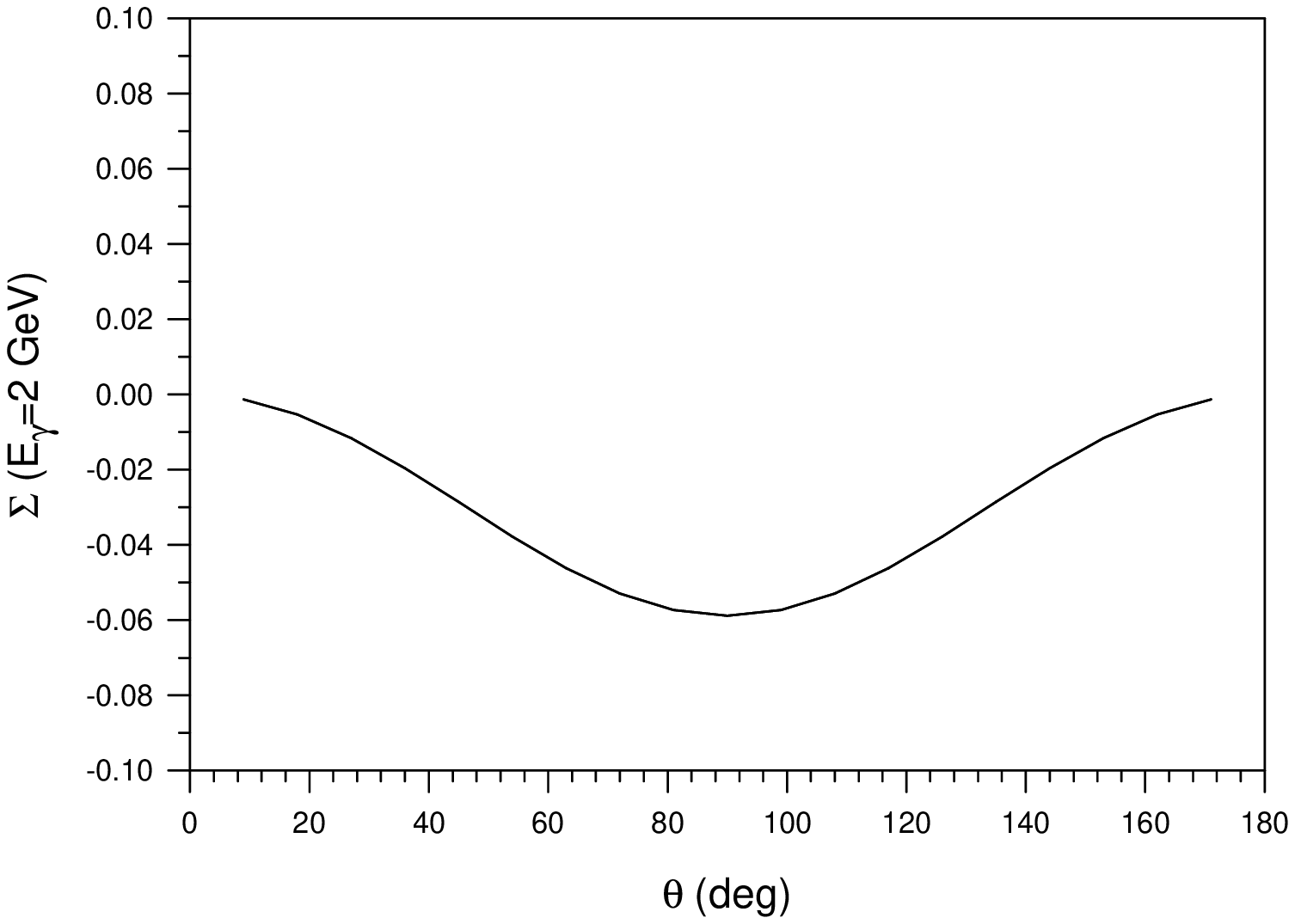} \\
(a) & (b)
\end{tabular}
\end{center}
\caption{\label{fig:Sigma} Same as Fig.~\ref{fig:Py} but
for the asymmetry $\Sigma$.  The data points are
from \protect\cite{SigmaData}.
}
\end{figure}

With these spinors as input, the amplitudes $\epsilon_\nu M_X^\nu$ can be evaluated
in terms of Dirac matrix and spinor products and then combined to construct
the predefined amplitudes $F_{i\pm}$.
At large $s$, these RNHA predictions for the helicity amplitudes reduce to
\bea
F_{1+}&\sim& 4\frac{\sqrt{2}}{\sqrt{s}}{\rm csc}^2(\frac{\theta_p}{2})G_{En}(\theta_p)G_{Ep}(\theta_p), \;\;
F_{1-}\sim 0, \\
F_{2+}&\sim& 2\frac{m}{s}
        \cot^3(\frac{\theta_p}{2})[G_{En}(\theta_p)G_{Mp}(\theta_p) - 
            G_{Mn}(\theta_p)G_{Ep}(\theta_p)], \\
F_{2-}&\sim& 2\frac{m}{s}
        \cot(\frac{\theta_p}{2})[G_{Mn}(\theta_p)G_{Ep}(\theta_p) - 
            G_{En}(\theta_p)G_{Mp}(\theta_p)], \nonumber \\
F_{3+}&\sim& 0, \;\;
F_{3-}\sim 0, \\
F_{4+}&\sim& -4\sqrt{2}\frac{m}{s}\cot(\frac{\theta_p}{2})G_{Mn}(\theta_p)G_{Ep}(\theta_p), \;\;
F_{4-}\sim 4\sqrt{2}\frac{m}{s}\cot(\frac{\theta_p}{2})G_{En}(\theta_p)G_{Mp}(\theta_p), \\
F_{5+}&\sim& \frac{2}{\sqrt{s}}\cot^2(\frac{\theta_p}{2})G_{En}(\theta_p)G_{Ep}(\theta_p), \;\;
F_{5-}\sim \frac{2}{\sqrt{s}}G_{En}(\theta_p)G_{Ep}(\theta_p), \\
F_{6+}&\sim& 4\sqrt{2}\frac{m}{s}\cot^3(\frac{\theta_p}{2})G_{En}(\theta_p)G_{Mp}(\theta_p), \;\;
F_{6-}\sim -4\sqrt{2}\frac{m}{s}\cot(\frac{\theta_p}{2})G_{Mn}(\theta_p)G_{Ep}(\theta_p).
\eea
From these we can calculate the various observables.  Plots of the results
and recent data~\cite{EgammaData,AngleData,SigmaData}
are given in Figs.~\ref{fig:Py}, \ref{fig:Cxprime}, \ref{fig:Czprime},
and \ref{fig:Sigma}.  Because the tree-level amplitudes are real, $P_y$ is automatically zero. 
That $C_{x'}$ is of order $m/\sqrt{s}$, rather than zero, is a correction to
hadron helicity conservation~\cite{HHC}.  Also, we find the asymmetry $\Sigma(90^\circ)$ 
to be approximately -0.06, rather than the nominal expectation~\cite{NominalSigma} of -1.
In general, the trends with photon energy seem to be modestly consistent with data.

\section{Electrodisintegration}
\label{sec:electrodis}

\begin{figure}[ht]
\vspace{0.2in}
\centerline{\includegraphics[width=12cm]{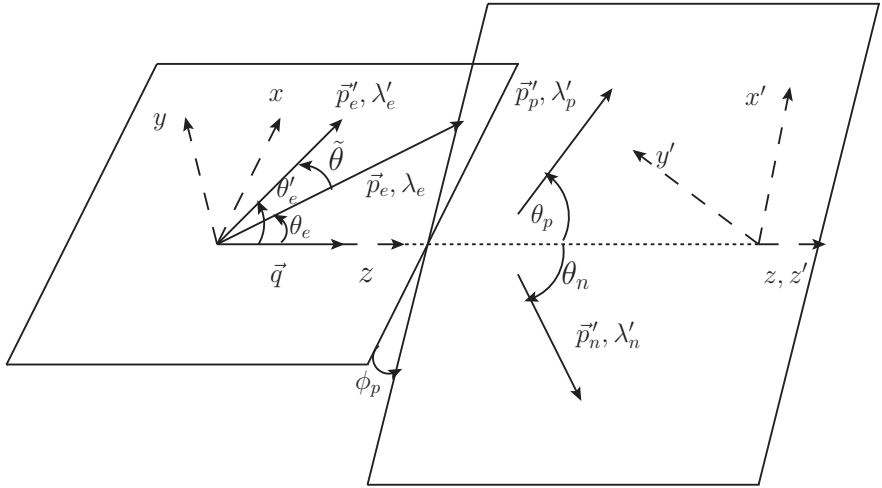}}
\caption{\label{fig:electrodis} Kinematics for deuteron electrodisintegration.
The unprimed axes are defined relative to the electron scattering plane,
and the primed axes relative to the final nucleon momenta.  The final proton
momentum has polar angle $\theta_p$ and azimuthal angle $\phi_p$ relative to
the unprimed frame.
}
\end{figure}

The kinematics of the electrodisintegration process are shown
in Fig.~\ref{fig:electrodis}.  The initial (final) momentum
and helicity of the electron are $p_e$ ($p'_e$) and $\lambda_e$ ($\lambda'_e$).
The intermediate photon carries four-momentum $q$.
The azimuthal angle $\phi_p$ of the proton measures the rotation of
the hadronic reaction plane relative to the electron 
scattering plane.  

In the lab frame, with the $z$ axis taken along the photon three-momentum and the electron
mass neglected, the initial and final electron four-momenta are
\be
p_e=(E_e,E_e\sin\theta_e,0,E_e\cos\theta_e),\;\;
p'_e=(E'_e,E'_e\sin\theta'_e,0,E'_e\cos\theta'_e),
\ee
with $E'_e$ and $\tilde\theta=\theta'_e-\theta_e$,
the angle of the scattered electron to the beam direction, being measured.
The photon four-momentum $q=(E_\gamma,0,0,q_z)$ is just $p_e-p'_e$, which yields
\be
Q^2\equiv -q^2=2 E_e E'_e(1-\cos\tilde\theta),\;\;
E_\gamma=E_e-E'_e,\;\;
q_z=\sqrt{E_\gamma^2+Q^2}.
\ee
The deuteron four-momentum is $p=(m_d=2m,0,0,0)$, and in the
zero-binding limit, the initial proton and neutron four-momenta 
are $p_p=p_n=(m,0,0,0)$.  The final nucleon four-momenta are
\be
p'_p=(E'_p=\sqrt{\vec{p}^{\,\prime 2}_p+m^2},
      |\vec{p}^{\,\prime}_p|\sin\theta_p\cos\phi_p,
      |\vec{p}^{\,\prime}_p|\sin\theta_p\sin\phi_p,
      |\vec{p}^{\,\prime}_p|\cos\theta_p),
\ee
\be
p'_n=(E'_n=\sqrt{\vec{p}^{\,\prime 2}_n+m^2},
      -|\vec{p}^{\,\prime}_n|\sin\theta_n\cos\phi_p,
      -|\vec{p}^{\,\prime}_n|\sin\theta_n\sin\phi_p,
      |\vec{p}^{\,\prime}_n|\cos\theta_n).
\ee

Within the one-photon-exchange approximation,
the scattering amplitude is proportional to
\be \label{eq:Med}
M_{ed}(\lambda'_p,\lambda'_n,\lambda'_e;\lambda_d,\lambda_e)
=\bar{u}'_e\gamma_\mu u_e\frac{D_{\mu\nu}}{q^2}M^\nu(\lambda'_p,\lambda'_n,\lambda_d),
\ee
with $u_e$ ($u'_e$) the initial (final) spinor of the electron
and $M^\nu$ given in (\ref{eq:Mnu}).  The numerator of the photon progator
is the sum over photon polarizations
\be  \label{eq:polarizationsum}
D_{\mu\nu}=\sum_{\lambda=-1}^1 (-1)^\lambda \epsilon_\mu^*(\lambda)\epsilon_\nu(\lambda).
\ee
The polarization four-vectors are\footnote{In the hadronic c.m.\ frame,
the longitudinal polarization vector is $\epsilon(0)=(q'_z/Q,0,0,E'_\gamma/Q)$.}
\be \label{eq:polarizationvectors}
\epsilon(\pm1)=\mp\frac{1}{\sqrt{2}}(0,1,\pm i,0), \;\;
\epsilon(0)=(q_z/Q,0,0,E_\gamma/Q)
\ee
relative to the photon four-momentum $q=(E_\gamma,0,0,q_z)$.
Polarization observables~\cite{Jeschonnek1,Jeschonnek2a,Jeschonnek2b,Jeschonnek3,%
Laget,Atti,Sargsian2009,Arenhovel,Gakh,Raskin,Dmitrasinovic}
can then be computed from these helicity amplitudes.

In keeping with the notation of \cite{Jeschonnek2a,Jeschonnek2b} and \cite{Dmitrasinovic},
the differential cross section for electrodisintegration, summed over the
final electron and neutron helicities in the lab
frame, is~\cite{Raskin,Dmitrasinovic}\footnote{In \protect\cite{Jeschonnek2a}, 
$h$ is $2\lambda_e$ but in \protect\cite{Dmitrasinovic}, $h$ is just $\lambda_e$,
which leads to additional factors of 2.}
\bea
d\sigma&\equiv&\frac{d\sigma^5}{dE' d\Omega'_e d\Omega'_p} \\
&=&\frac{m_p m_n |\vec{p}^{\,\prime}_p|}{16\pi^3 m_d} \frac{\sigma_{\rm Mott}}{f_{\rm rec}}
\left[\nu_LR_L+\nu_TR_T+\nu_{TT}R_{TT}+\nu_{LT}R_{LT}
        +2\lambda_e\nu_{LT'}T_{LT'}+2\lambda_e\nu_{T'}R_{T'}\right], \nonumber
\eea
where $\Omega'_e$ ($\Omega'_p$) is the solid angle of the scattered electron
(proton), $\sigma_{\rm Mott}$ is the Mott cross section,
$f_{\rm rec}=|1+(E_\gamma |\vec{p}^{\,\prime}_p|-E'_p q_z\cos\theta_p)/(m_d |\vec{p}^{\,\prime}_p|)|$
is the lab recoil factor,
\bea
\nu_L&=&\frac{Q^4}{q_z^4},\;\;
\nu_T=\frac{Q^2}{2q_z^2}+\tan^2\frac{\tilde\theta}{2}, \;\;
\nu_{TT}=\frac{Q^2}{2q_z^2}, \;\;
\nu_{LT}=\frac{Q^2}{\sqrt{2}q_z^2}\sqrt{\frac{Q^2}{q_z^2}+\tan^2\frac{\tilde\theta}{2}}, \\
\nu_{LT'}&=&-\frac{Q^2}{\sqrt{2}q_z^2}\tan\frac{\tilde\theta}{2},\;\;
\nu_{T'}=\tan\frac{\tilde\theta}{2}\sqrt{\frac{Q^2}{q_z^2}+\tan^2\frac{\tilde\theta}{2}},
\nonumber
\eea
and $\tilde\theta=\theta'_e-\theta_e$ is the angle between the incoming and outgoing electron.
The response functions $R_X$ depend upon the hadronic
helicity amplitudes and the azimuthal angle $\phi_p$ of the hadronic
scattering plane.  The subscripts refer to the polarization of the
intermediate photon, which enters on substitution of the polarization
expansion (\ref{eq:polarizationsum})
for the numerator of the photon propagator in the hadronic amplitude (\ref{eq:Med}).
The amplitude then decomposes into separate leptonic and hadronic factors
\be
M_{ed}(\lambda'_p,\lambda'_n,\lambda'_e;\lambda_d,\lambda_e)
=-\sum_{\lambda=-1}^1 \bar{u}'_e\fsl{\epsilon}^*(\lambda) u_e \frac{(-1)^\lambda}{Q^2}
         \epsilon_\nu(\lambda)M^\nu(\lambda'_p,\lambda'_n,\lambda_d).
\ee
The leptonic factors give rise to the $\nu_X$ coefficients,
and the hadronic factors to the response functions in the square of 
the amplitude used to construct the cross section~\cite{Dmitrasinovic}.
The subscript L(T) indicates a purely longitudinal (transverse)
contribution, while LT is a cross term between longitudinal and
transverse photon helicities.  The TT subscript marks a cross term
between different transverse helicities.  A prime indicates a 
different combination of transverse helicities.

The response functions are computed from components of the hadronic tensor
\be
w_{\lambda',\lambda}=\frac23\sum_{\lambda''_p,\lambda'_p,\lambda'_n,\lambda''_d,\lambda_d}
   \epsilon_\nu^*(\lambda')M^{\nu*}(\lambda''_p,\lambda'_n,\lambda''_d)\rho^p_{\lambda''_p,\lambda'_p}
   \epsilon_\mu(\lambda)M^\mu(\lambda'_p,\lambda'_n,\lambda_d)\rho^d_{\lambda''_d,\lambda_d},
\ee
with $\rho^p$($\rho^d$) the density matrix for the proton (deuteron) helicity state.
We construct these in the $xyz$ coordinate system of the electron scattering
plane.  The particular components are~\cite{Jeschonnek2a}
\bea \label{eq:responsefunctions}
R_L&=&w_{0,0}, \;\; 
R_T=w_{1,1}+w_{-1,-1}, \;\; 
R_{T'}=w_{1,1}-w_{-1,-1}, \\
R_{TT}&=&2 {\rm Re}w_{1,-1}, \;\;
R_{LT}=-2 {\rm Re}\left[w_{0,1}-w_{0,-1}\right], \;\;
R_{LT'}=-2 {\rm Re}\left[w_{0,1}+w_{0,-1}\right]. \nonumber
\eea

For an unpolarized target, the deuteron density matrix is
proportional to the identity, $\rho^d=\frac13 I$; similarly,
if the proton helicity is not detected, $\rho^p=\frac12 I$.
We then have the unpolarized cross section~\cite{Jeschonnek2a}
\be \label{eq:unpolsigma}
d\sigma_{\rm unpol}
=\frac{m_p m_n |\vec{p}^{\,\prime}_p|}{16\pi^3 m_d} \frac{\sigma_{\rm Mott}}{f_{\rm rec}}\sigma_0,\;\;
\sigma_0\equiv\nu_LR_L^U+\nu_TR_T^U+\nu_{TT}R_{TT}^U+\nu_{LT}R_{LT}^U,
\ee
where the $R_X^U$ are computed with the simple density matrices.
These are then computable in our model, with the basic computation being the
evaluation of $\epsilon(\lambda_\gamma)_\mu M^\mu$, which differs from the 
photodisintegration calculation in only two ways: $Q^2$ is not zero
and $\lambda_\gamma$ ranges over all three possibilities.  

The unpolarized response functions $R_{LT'}^U$ and $R_{T'}^U$ are identically zero.
With $\rho^d$ replaced by $\frac13 I$ and the form (\ref{eq:polarizationvectors})
of the polarization vectors taken into account, ${\rm Re}w(0,1)$ is just the negative of
${\rm Re}w(0,-1)$, and $w_{1,1}$ is equal to $w_{-1,-1}$.  Thus, the inputs to $R_{LT'}^U$ 
and $R_{T'}^U$, as given in (\ref{eq:responsefunctions}), immediately cancel.

The recent $ed\rightarrow e'pn$ experiment at JLab~\cite{Yero} does not
include polarization but does begin to reach momentum transfers
sufficient to consider the RNHA approach. Once polarization data is
available, the expressions developed here and in the Appendix can be
compared.

\section{Summary}
\label{sec:summary}

We have extended the reduced nuclear amplitude approach~\cite{BrodskyChertok,RNA}
to helicity amplitudes and applied this model to analysis
of elastic electron-deuteron scattering, deuteron photodisintegration,
and deuteron electrodisintegration.  These are just examples of the
approach, which is generally applicable to exclusive nuclear
processes.  The primary limitation is that, for any process, the 
net momentum transfer to every nucleon must be large; therefore,
as the number of nucleons increases, the required 
beam energy can increase dramatically.  The primary gain
is precocious scaling in the dependence on momentum 
transfer.  What the model (or the original RNA approach) does
not provide, though, is an overall normalization; comparisons
must be made in terms of ratios.

By considering helicity amplitudes, many more quantities
can be studied, including polarization dependence. All
three of the deuteron's electromagnetic form factors can
be calculated and from there various elastic scattering
observables can be constructed.  In Sec.~\ref{sec:elasticscattering}
we considered the standard structure functions $A$ and $B$ as 
well as the tensor polarizations $t_{2m}$.  Generally, the
model implies the need for momentum transfers larger
than one would have hoped for seeing simple perturbative
QCD scaling.  However, our results do imply that
the deuteron structure function $B$ is a good
place to look, above a transfer of 10 GeV$^2$.

The RNHA results for polarization observables in deuteron
photodisintegration, considered in Sec.~\ref{sec:photodis},
are somewhat consistent with experiment.  In particular,
our result for the asymmetry $\Sigma$, with a value of 
$\Sigma(90^\circ)\simeq -0.06$, is much better than
the value of -1 originally expected~\cite{NominalSigma}.
Higher photon energies would, of course, be useful.

We have also constructed the RNHA framework for analysis
of deuteron electrodisintegration, in Sec.~\ref{sec:electrodis}.
This stands ready for comparison with experiment
when data is available at sufficient energies.
One aspect that does remain is to consider
polarization of the outgoing proton, in addition
to polarization of the beam and target.

Other processes that one might consider include
deeply virtual Compton scattering on the deuteron,
pion photoproduction on the deuteron~\cite{pion},
and photodisintegration of $^3$He~\cite{3He}.
In each case, our approach can provide not only
information about helicity amplitudes but also
an analysis of nonleading momentum transfer dependence
with respect to the onset of perturbative QCD scaling.
We look forward to experiments at larger momentum
transfers for all of these processes.

\acknowledgments
This work began in conversations with S.J. Brodsky and D.-S. Hwang.
Some calculations were checked by W. Miller and C. Salveson.
Diagrams were drawn with use of JaxoDraw~\cite{JaxoDraw}.

\appendix

\section{Electrodisintegration with polarization}  \label{sec:appendix}

If we consider polarization for the beam and the target,\footnote{%
For discussion of a polarized outgoing proton, see \protect\cite{Jeschonnek2b}
and \protect\cite{Dmitrasinovic}.}
the proton density matrix is still just $\rho^p=\frac12 I$, but the deuteron
density matrix in the $xyz$ frame is~\cite{Jeschonnek2a}
\be
\rho^d=\frac13\left(\begin{array}{ccc} 
  1+\sqrt{\frac32}T_{10}+\frac{1}{\sqrt{2}}T_{20} & -\sqrt{\frac32}(T_{11}^*+T_{21}^*) & \sqrt{3}T_{22}^* \\
  -\sqrt{\frac32}(T_{11}+T_{21}) & 1-\sqrt{2}T_{20} & -\sqrt{\frac32}(T_{11}^*-T_{21}^*) \\
  \sqrt{3}T_{22} & -\sqrt{\frac32}(T_{11}-T_{21}) & 1-\sqrt{\frac32}T_{10}+\frac{1}{\sqrt{2}}T_{20}
  \end{array}\right).
\ee
For a target polarization defined relative to the beam direction, rather than
the $xyz$ system used above, the tensor polarization coefficients $T_{JM}$
are related to the coefficients $\tilde{T}_{JM}$ defined relative to the beam~\cite{Jeschonnek2a}.
If only $\tilde{T}_{10}$ and $\tilde{T}_{20}$ are nonzero,\footnote{The spherical
tensor moments are related to the Cartesian tensor moments as $\tilde{T}_{10}=\sqrt{\frac32}P_z$
and $\tilde{T}_{20}=\frac{1}{\sqrt{2}}P_{zz}$.} the nonzero $T_{JM}$ are 
\bea
T_{10}&=&\cos\tilde\theta\,\tilde{T}_{10}, \;\;
T_{11}=-\frac{1}{\sqrt{2}}\sin\tilde\theta\,\tilde{T}_{10}, \\ 
T_{20}&=&\frac14(1+3\cos 2\tilde\theta)\tilde{T}_{20},   \;\;
T_{21}=-\sqrt{\frac38}\sin 2\tilde\theta\,\tilde{T}_{20}, \;\;
T_{22}=\sqrt{\frac{3}{32}}(1-\cos 2\tilde\theta)\tilde{T}_{20}. \nonumber
\eea
The density matrix can then be written as
\be
\rho^d=\left(\frac13I+\tilde{T}_{10}\rho^{dV}+\tilde{T}_{20}\rho^{dT}\right),
\ee
where
\be
\rho^{dV}=\frac13\left(\begin{array}{ccc} 
  \sqrt{\frac32}\cos\tilde\theta & \frac{\sqrt{3}}{2}\sin\tilde\theta & 0 \\
  \frac{\sqrt{3}}{2}\sin\tilde\theta & 0 & \frac{\sqrt{3}}{2}\sin\tilde\theta \\
  0 & \frac{\sqrt{3}}{2}\sin\tilde\theta & -\sqrt{\frac32}\cos\tilde\theta
  \end{array}\right)
\ee
and
\be
\rho^{dT}=\frac13\left(\begin{array}{ccc} 
 \frac{1}{4\sqrt{2}}(1+3\cos 2\tilde\theta) & \frac34\sin 2\tilde\theta & \frac{3}{\sqrt{32}}(1-\cos 2\tilde\theta) \\
  \frac34\sin 2\tilde\theta & -\frac{1}{2\sqrt{2}}(1+3\cos 2\tilde\theta) & -\frac34\sin 2\tilde\theta \\
  \frac{3}{\sqrt{32}}(1-\cos 2\tilde\theta)& -\frac34\sin 2\tilde\theta & \frac{1}{4\sqrt{2}}(1+3\cos 2\tilde\theta)
  \end{array}\right).
\ee
The response functions can then be separated into unpolarized, vector, and tensor contributions
as $R_X=R_X^U+\tilde{T}_{10}R_X^V+\tilde{T}_{20}R_X^T$, with $R_X^U$, $R_X^V$, and $R_X^T$ computed
with $\rho^d$ replaced by $\frac13 I$, $\rho^{dV}$, and $\rho^{dT}$, respectively.  

With $d\sigma_{\rm unpol}$ defined as the unpolarized cross section, given in (\ref{eq:unpolsigma}),
the full cross section can be written as
\be
d\sigma=\left[1+\tilde{T}_{10} \left(A_d^V +2\lambda_e A_{ed}^V\right)
  +\tilde{T}_{20}\left(A_d^T+2\lambda_e A_{ed}^T\right)\right]d\sigma_{\rm unpol},
\ee
in terms of the single and double asymmetries
\bea
A_d^V&=&\left[\nu_L R_L^V+\nu_T R_T^V+\nu_{TT} R_{TT}^V+\nu_{LT} R_{LT}^V\right]/\sigma_0, \\
A_{ed}^V&=&\left[\nu_{LT'} R_{LT'}^V+\nu_{T'} R_{T'}^V\right]/\sigma_0, \\
A_d^T&=&\left[\nu_L R_L^T+\nu_T R_T^T+\nu_{TT} R_{TT}^T+\nu_{LT} R_{LT}^T\right]/\sigma_0, \\
A_{ed}^T&=&\left[\nu_{LT'} R_{LT'}^T+\nu_{T'} R_{T'}^T\right]/\sigma_0.
\eea
For a recent summary of data, see \cite{Mayer}.


\end{document}